\documentclass[12pt]{article}

\usepackage{latexsym}
\usepackage{verbatim}
\usepackage{amssymb}

\catcode`\@=11
\def\marginnote#1{}
\newcount\hour
\newcount\minute
\newtoks\amorpm
\hour=\time\divide\hour by60
\minute=\time{\multiply\hour by60 \global\advance\minute by-\hour}
\edef\standardtime{{\ifnum\hour<12 \global\amorpm={am}%
        \else\global\amorpm={pm}\advance\hour by-12 \fi
        \ifnum\hour=0 \hour=12 \fi
        \number\hour:\ifnum\minute<10 0\fi\number\minute\the\amorpm}}
\edef\militarytime{\number\hour:\ifnum\minute<10 0\fi\number\minute}
\def\draftlabel#1{{\@bsphack\if@filesw {\let\thepage\relax
   \xdef\@gtempa{\write\@auxout{\string
      \newlabel{#1}{{\@currentlabel}{\thepage}}}}}\@gtempa
   \if@nobreak \ifvmode\nobreak\fi\fi\fi\@esphack}
        \gdef\@eqnlabel{#1}}
\def\@eqnlabel{}
\def\@vacuum{}
\def\draftmarginnote#1{\marginpar{\raggedright\scriptsize\tt#1}}
\def\draft{\oddsidemargin -.5truein
        \def\@oddfoot{\sl preliminary draft \hfil
        \rm\thepage\hfil\sl\today\quad\militarytime}
        \let\@evenfoot\@oddfoot \overfullrule 3pt
        \let\label=\draftlabel
        \let\marginnote=\draftmarginnote
   \def\@eqnnum{(\theequation)\rlap{\kern\marginparsep\tt\@eqnlabel}%
\global\let\@eqnlabel\@vacuum}  }

\def\preprint{\twocolumn\sloppy\flushbottom\parindent 1em
        \leftmargini 2em\leftmarginv .5em\leftmarginvi .5em
        \oddsidemargin -.5in    \evensidemargin -.5in
        \columnsep 15mm \footheight 0pt
        \textwidth 250mmin      \topmargin  -.4in
        \headheight 12pt \topskip .4in
        \textheight 175mm
        \footskip 0pt
        \def\@oddhead{\thepage\hfil\addtocounter{page}{1}\thepage}
        \let\@evenhead\@oddhead \def\@oddfoot{} \def\@evenfoot{} }

\def\titlepage{\@restonecolfalse\if@twocolumn\@restonecoltrue\onecolumn
     \else \newpage \fi \thispagestyle{empty}\c@page\z@ 
        \def\thefootnote{\fnsymbol{footnote}} }

\def\endtitlepage{\if@restonecol\twocolumn \else  \fi
        \def\thefootnote{\arabic{footnote}}
        \setcounter{footnote}{0}}  %\c@footnote\z@ }

\catcode`@=12
\relax
\def\bea{\begin{array}}
\def\bem{\begin{displaymath}}
\def\beq{\begin{equation}}

\def\eea{\end{array}}
\def\eem{\end{displaymath}}
\def\eeq{\end{equation}}
\def\Im{\mathop{\rm Im}}

\def\ov{\overline}

\def\Re{\mathop{\rm Re}}

\def\s2w{\sin^2 \theta_W}

\relax
\def\dalpha{{\dot\alpha}}
\def\dbeta{{\dot\beta}}

\def\crbig{\\\noalign{\vspace {3mm}}}
\def\bigint{{\displaystyle\int}}

\def\Dint{\bigint d^2\theta d^2\ov\theta\,}
\def\Fint{\bigint d^2\theta\,}
\def\Fbarint{\bigint d^2\ov\theta\,}
\relax
\begin{document}
\topmargin-2.4cm
%\draft
%\preprint
%
%
%
%
\begin{titlepage}
\begin{flushright}
CERN-PH-TH/2009-232 \\
November 2009
\end{flushright}
\vspace{1.0cm}

\begin{center}{\Large\bf
Nonlinear Supersymmetry, Brane-bulk Interactions \\ \vspace{3mm}
and Super-Higgs without Gravity}
\vspace{1.0cm}

{\large\bf N. Ambrosetti$^1$, I. Antoniadis$^{2,3}$, J.-P. Derendinger$^{1,3}$
\\ 
\vspace{3mm}
and P. Tziveloglou$^{2,4}$}

\vspace{6mm}

{\small 
$^1$ Albert Einstein Center for Fundamental Physics \\
Institute for Theoretical Physics, Bern University \\
Sidlerstrasse 5, CH--3012 Bern, Switzerland \\

\vspace{2mm}

$^2$ Department of Physics, CERN - Theory Division \\
CH--1211 Geneva 23, Switzerland \\

\vspace{2mm}

$^3$ Centre de Physique Th\'eorique, UMR du CNRS 7644 \\
Ecole Polytechnique, F--91128 Palaiseau, France

\vspace{2mm}

$^4$ Department of Physics, Cornell University, Ithaca, NY 14853, USA 
}

\end{center}
\vspace{.6cm}

\begin{center}
{\large\bf Abstract}
\end{center}
\begin{quote}
We derive the coupling of a hypermultiplet of $N=2$ global supersymmetry to the Dirac-Born-Infeld
Maxwell theory with linear $N=1$ and a second nonlinear supersymmetry. At the level of global
supersymmetry, this construction corresponds to the interaction with Maxwell brane fields of bulk
hypermultiplets, such as the 
universal dilaton of type IIB strings compactified on a Calabi-Yau manifold.
It displays in particular the
active role of a four-form field. Constrained $N=1$ and $N=2$ superfields and the formulation of the hypermultiplet in its single-tensor version are used to derive the nonlinear realization, allowing a 
fully off-shell description. Exact results with explicit
symmetries and supersymmetries are then obtained. The 
electric-magnetic dual version of the theory is also derived and the gauge structure of the
interaction is exemplified with $N=2$ nonlinear QED of a charged hypermultiplet.
Its Higgs phase describes a novel super-Higgs mechanism without gravity, where the goldstino is 
combined with half of the hypermultiplet into an $N=1$ massive vector multiplet.

\end{quote}

\end{titlepage}
\renewcommand{\theequation}{\arabic{section}.\arabic{equation}}

\setcounter{footnote}{0}
\setcounter{page}{0}
\setlength{\baselineskip}{.6cm}
\setlength{\parskip}{.2cm}
\newpage
%
% BODY
%

\newpage

%***************************************
\section{Introduction} \label{secintro}
%***************************************
\setcounter{equation}{0}

It is notorious that (linear) $N=2$ supersymmetry, global or local, forbids a dependence 
on hypermultiplet scalars of gauge kinetic terms. For instance, in $N=2$ supergravity, the scalar 
manifold is the product of a quaternion-K\"ahler (Einstein) manifold, for hypermultiplet 
scalars \cite{BW}, and a K\"ahler manifold of a special type for vector multiplet scalars 
\cite{dWLVP}. In global $N=2$ supersymmetry, the quaternion-K\"ahler manifold of hypermultiplet
scalars is replaced by a Ricci-flat hyperk\"ahler space \cite{AGF}.

If however (at least) one of the supersymmetries is nonlinearly realized, these restrictions on the action 
are expected to change. For instance, string theory indicates that the 
Dirac-Born-Infeld (DBI) Lagrangian describing kinetic terms of brane gauge fields may interact with 
the dilaton and with its hypermultiplet partners. Moreover, if the dilaton supermultiplet 
is formulated with one or two antisymmetric tensors, more involved interactions dictated by the gauge 
symmetries of the theory are certainly allowed. An interesting problem is then to construct an 
interaction Lagrangian in which, when the second supersymmetry turns nonlinear,
both the DBI Lagrangian and its necessary dilaton dependence are simultaneously generated.
In other words, if we consider a theory with a broken, nonlinear supersymmetry realized in a 
goldstino mode, another unbroken linear
supersymmetry and a DBI super-Maxwell system coupled to hypermultiplet fields, we certainly 
expect that the allowed Lagrangians are severely restricted.
Analyzing these restrictions is the main motivation of this paper.

In this work, we construct an action invariant under $N=2$ global supersymmetry, one of them 
being nonlinearly realized, involving the Maxwell goldstino multiplet of the nonlinear supersymmetry 
coupled to a single-tensor $N=2$ multiplet \cite{dWvH, LR, Ketal}, or equivalently to a hypermultiplet
with one abelian (shift) isometry. In the absence of this multiplet, the action              
reduces to the standard super-Maxwell DBI theory, derived in the past from the same symmetry 
principle \cite{BG, RT, ADM}. The coupling of the two multiplets is shown to arise from a $N=2$
Chern-Simons (CS) term which, under electric-magnetic duality, amounts to shifting the gauge field strength by the antisymmetric tensor. Moreover, under Poincar\'e duality of the antisymmetric tensor 
to a pseudoscalar, the CS coupling becomes a St\"uckelberg gauging of the
pseudoscalar axionic symmetry. 

An important property of the single-tensor multiplet is that it admits an off-shell (superspace) 
formulation, unlike the generic hypermultiplet that can be formulated off-shell only at the cost of 
introducing infinite number of auxiliary fields in the context of harmonic superspace \cite{harm}. 
Thus, our formalism using the single-tensor $N=2$ multiplet allows to construct off-shell 
supersymmetric Lagrangians. By an appropriate change of variables from the $N=2$ 
single-tensor multiplet, one finds an action that couples the goldstino vector multiplet (of the 
linear supersymmetry) to an $N=2$ charged hypermultiplet, describing the low-energy limit 
of a theory with partial spontaneous supersymmetry breaking from $N=2$ to $N=1$~\cite{APT, Itoyama}. 

The vacuum of this theory exhibits an interesting novel feature: the goldstino is `absorbed' into
a massive vector multiplet of $N=1$ linear supersymmetry, leaving a massless $N=1$ 
chiral multiplet associated to flat directions of the scalar potential. The goldstino assembles
with one of the two Weyl fermions in the single-tensor multiplet to form a massive Dirac spinor.
At one particular point along the flat directions, the vector multiplet becomes massless and the 
$U(1)$ is restored. This phenomenon is known from D-brane dynamics, where the $U(1)$ 
world-volume field becomes generically massive due to the CS coupling. 
A crucial role for the invariance of the action under nonlinear 
supersymmetry is played by a non-dynamical four-form gauge potential, known again from 
D-brane dynamics. Hence, a globally supersymmetric combination of Higgs and super-Higgs 
mechanisms, in the presence of a four-form field, eliminates any massless goldstino fermion related
to partial supersymmetry breaking. This interesting new mechanism can be studied in the context 
of nonlinear $N=2$ quantum electrodynamics with one charged hypermultiplet, which after
a holomorphic field redefinition and a duality transformation, is equivalent to our setup.

In type IIB superstrings compactified to four dimensions with eight residual supercharges, the dilaton 
scalar (associated to the string coupling) belongs to a universal hypermultiplet, together with the 
(Neveu-Schwarz) NS--NS antisymmetric tensor and the (Ramond) R--R scalar and two-form. Its 
natural basis is therefore a double-tensor supermultiplet,\footnote{This representation of $N=2$ 
global supersymmetry has been only recently explicitly constructed \cite{DTmult}. See also 
Ref.~\cite{TV}.} having three perturbative 
isometries associated to the two axionic shifts of the antisymmetric tensors and an extra shift of the 
R--R scalar. These isometries form a Heisenberg algebra, which at the string tree-level is enhanced 
to the quaternion-K\"ahler and K\"ahler space $SU(2,1)/SU(2)\times U(1)$.
At the level of global $N=2$, imposing the Heisenberg algebra of isometries determines a unique
hyperk\"ahler manifold of dimension four, depending on a single parameter, in close analogy with
the local case of a quaternionic space where the corresponding parameter is associated
to the one-loop correction~\cite{oneloop}. 
This manifold is {\em not} trivially flat and should describe the rigid limit of the universal 
hypermultiplet.

The plan of the paper is as follows. In Section \ref{seclinear}, we review the construction of the $N=2$ 
simple-tensor and Maxwell supermultiplets in terms of $N=1$ superfields and we describe
their interaction in a Chern-Simons term, as was earlier partly done in Ref.~\cite{ADM}. In addition we 
explain how the intricate web of gauge variations in the St\"uckelberg coupling of the Maxwell and 
single-tensor supermultiplets leads to the interpretation of one (non-propagating) component of the 
single-tensor as a four-form field. In Section \ref{secchiralN=2}, we reformulate the supermultiplets
in chiral $N=2$ superspace and then demonstrate how this construction can be used to describe 
electric-magnetic duality in a manifestly $N=2$ covariant way. In Section \ref{secDBI}, 
we first review the construction of the Dirac-Born-Infeld theory from constrained 
$N=2$ superfields describing the goldstino of one 
non-linear supersymmetry and then extend it to construct its coupling to a single-tensor 
supermultiplet, engineered by a CS term. We also perform an electromagnetic duality to determine 
the `magnetic' version of the theory. WIth the dilaton 
hypermultiplet of type IIB superstrings in mind, we impose the Heisenberg algebra of perturbative 
isometries to our theory. In Section \ref{secQED}, we derive the coupling of the 
Maxwell goldstino multiplet to a charged hypermultiplet and make a detailed analysis of the
vacuum structure of $N=2$ super-QED with partial supersymmetry breaking. We conclude in 
Section \ref{secfinal} and two appendices present our conventions and the resolution of a 
quadratic constraint applied on a $N=2$ chiral superfield.

%***************************************************************************
\section{The linear $\bf{N=2}$ Maxwell-dilaton system}\label{seclinear}
\setcounter{equation}{0}
%****************************************************************************

Our first objective is to describe, in the context of linear $N=2$ supersymmetry, 
the coupling of the single-tensor multiplet to $N=2$ super-Maxwell theory. Since these
two supermultiplets admit off-shell realizations, they can be described in superspace without reference 
to a particular Lagrangian.  Gauge transformations of the Maxwell multiplet use a single-tensor
multiplet, we then begin with the latter.

%***************************************************
\subsection{The single-tensor multiplet}\label{secST}
%***************************************************

In global $N=1$ supersymmetry, a real antisymmetric tensor field $b_{\mu\nu}$ is described
by a chiral, spinorial superfield $\chi_\alpha$ with $8_B+8_F$ fields \cite{S}\footnote{The 
notation $m_B+n_F$
stands for `$m$ bosonic and $n$ fermionic fields'.}:
\beq
\label{ST2}
\chi_\alpha = - {1\over4} \theta_\alpha (C+iC^\prime) 
+ {1\over4}(\theta\sigma^\mu\ov\sigma^\nu)_\alpha \, b_{\mu\nu} + \ldots
\qquad\qquad
(\,\ov D_\dalpha\chi_\alpha=0\,),
\eeq
$C$ and $C^\prime$ being the real scalar partners of $b_{\mu\nu}$. 
The curl
$h_{\mu\nu\rho} = 3\, \partial_{[\mu} b_{\nu\rho]}$ is described by the real superfield
\beq
\label{ST1}
L=D^\alpha \chi_\alpha -\ov D_\dalpha \ov\chi^\dalpha. 
\eeq
Chirality of $\chi_\alpha$ implies linearity of $L$: $DDL=\ov{DD}L=0$. The linear superfield $L$
is invariant under the supersymmetric gauge transformation\footnote{$\Delta$ is an arbitrary real superfield.}
\beq
\label{ST3}
\chi_\alpha \quad\longrightarrow\quad \chi_\alpha + {i\over4}\ov{DD}D_\alpha \Delta, \qquad\qquad
\ov\chi_\dalpha \quad\longrightarrow\quad \ov\chi_\dalpha + {i\over4}DD \ov D_\dalpha \Delta,
\eeq
of $\chi_\alpha$: this is the supersymmetric extension of the invariance of $h_{\mu\nu\rho}$ under
$\delta b_{\mu\nu} = 2\,\partial_{[\mu}\Lambda_{\nu]}$. Considering bosons only, the gauge
transformation (\ref{ST3}) eliminates three of the six components of $b_{\mu\nu}$ and the scalar 
field $C^\prime$. Accordingly, $L$ only depends on the invariant curl $h_{\mu\nu\rho}$ and on the 
invariant real scalar $C$. The linear $L$ describes then $4_B+4_F$ fields.
Using either $\chi_\alpha$ or $L$, we will find two descriptions of the single-tensor multiplet of global
$N=2$ supersymmetry \cite{dWvH, LR, Ketal}.

In the gauge-invariant description using $L$, the $N=2$ multiplet is completed with a
chiral superfield $\Phi$ ($8_B+8_F$ fields in total). The second supersymmetry transformations (with parameter $\eta_\alpha$) are
\beq
\label{ST4}
\begin{array}{rcl}
\delta^* L &=& -\frac{i}{\sqrt 2} (\eta D\Phi +\ov{\eta D}\ov \Phi) \,, 
\crbig
\delta^* \Phi &=&  i \sqrt2 \, \ov{\eta D}L \,,\qquad\qquad \delta^* \ov\Phi \,\,=\,\, i \sqrt2 \,\eta D L \,,
\end{array}
\eeq
where $D_\alpha$ and $\ov D_\dalpha$ are the usual $N=1$ supersymmetry derivatives verifying
$\{ D_\alpha , \ov D_\dalpha \} = -2i (\sigma^\mu)_{\alpha\dalpha}\partial_\mu$. It is easily verified
that the $N=2$ supersymmetry algebra closes on $L$ and $\Phi$. 

We may try to replace $L$ by $\chi_\alpha$ with second supersymmetry transformation 
$\delta^*\chi_\alpha = -{i\over\sqrt2}\Phi\,\eta_\alpha$, as suggested when comparing Eqs.~(\ref{ST1})
and (\ref{ST4}). However, with superfields $\chi_\alpha$ and $\Phi$ only, the $N=2$ algebra only closes
up to a gauge transformation (\ref{ST3}). This fact, and the unusual number $12_B+12_F$
of fields, indicate that $(\chi_\alpha,\Phi)$ is a gauge-fixed version of the off-shell 
$N=2$ multiplet. We actually need another chiral $N=1$ superfield $Y$ to close the 
supersymmetry algebra. The second supersymmetry variations are
\beq
\label{ST5}
\begin{array}{rcl}
\delta^* Y &=& \sqrt2\, \eta\chi \, , 
\crbig
\delta^* \chi_\alpha &=& -{i\over\sqrt2} \Phi\,\eta_\alpha - {\sqrt2\over4} \eta_\alpha \, \ov{DD}\, \ov Y
-\sqrt2 i (\sigma^\mu\ov\eta)_\alpha \partial_\mu Y \, , 
\crbig
\delta^* \Phi &=&  2\sqrt2 i \left[\frac{1}{4}\,\ov{DD\eta\chi} 
+ i \partial_\mu\chi\sigma^\mu\ov\eta \right] .
\end{array}
\eeq
One easily verifies that the $Y$--dependent terms in $\delta^*\chi_\alpha$ induce a gauge transformation
(\ref{ST3}). Hence, the linear $L$ and its variation $\delta^*L$ do not feel $Y$.  
The superfields $\chi_\alpha$, $\Phi$ and $Y$ have $16_B+16_F$ field components. Gauge 
transformation (\ref{ST3}) eliminates $4_B+4_F$ fields. To further eliminate $4_B+4_F$ fields,
a new gauge variation 
\beq
\label{ST6}
Y \qquad\longrightarrow\qquad Y - {1\over2}\ov{DD} \Delta^\prime,
\eeq
with $\Delta^\prime$ real, is then postulated. We will see below that this variation is actually dictated by 
$N=2$ supersymmetry. There exists then a gauge in which $Y=0$ but in this gauge the 
supersymmetry algebra closes on $\chi_\alpha$ only up to a transformation (\ref{ST3}).
This is analogous to the Wess-Zumino gauge of $N=1$ supersymmetry, but in our case, 
this particular gauge respects $N=1$ supersymmetry and gauge symmetry (\ref{ST3}).

Two remarks should be made at this point. Firstly, the superfield $Y$ will play an important role in the
construction of the Dirac-Born-Infeld interaction with non-linear $N=2$ supersymmetry. As we will
see later on\footnote{See Subsection \ref{secV1XY}.}, it includes a four-index antisymmetric tensor field 
in its highest component. Secondly, a constant ($\theta$--independent) background value 
$\langle\Phi\rangle$ breaks the second supersymmetry only, $\delta^*\chi_\alpha = 
-{i\over\sqrt2}\langle\Phi\rangle \eta_\alpha+\ldots\,\,$ It is a natural source of partial supersymmetry 
breaking in the single-tensor multiplet. Notice that the condition $\delta^*\langle\Phi\rangle =0$
is equivalent to $\ov D_\dalpha(D\chi-\ov{D\chi})=0$.

An invariant kinetic action for the gauge-invariant single-tensor multiplet involves an arbitrary 
function solution of the 
three-dimensional Laplace equation (for the variables $L$, $\Phi$ and $\ov\Phi$) \cite{LR}:
\beq
\label{ST7}
{\cal L}_{ST} = \Dint {\cal H } (L, \Phi, \ov\Phi) \,, \qquad\qquad
{\partial^2{\cal H }\over\partial L^2} + 2 {\partial^2{\cal H }\over\partial\Phi\partial\ov\Phi} =0.
\eeq
In the dual hypermultiplet formulation the Laplace equation is replaced by a Monge-Amp\`ere equation.
We will often insist on theories with axionic shift symmetry $\delta\Phi = i c$ ($c$ real), dual to a 
double-tensor theory. In this case, ${\cal H}$ is a function of $L$ and $\Phi+\ov\Phi$ 
so that the general solution of Laplace equation is
\beq
\label{ST8}
{\cal L}_{ST} = \Dint \, H({\cal V}) + {\rm h. c.} ,
\qquad\qquad
{\cal V} = L + {i\over\sqrt2} (\Phi+\ov\Phi ),
\eeq
with an arbitrary analytic function $H({\cal V})$.

%****************************************************************************
\subsection{The Maxwell multiplet, Fayet-Iliopoulos terms} \label{secMaxwell}
%****************************************************************************

Take two real vector superfields $V_1$ and $V_2$. Variations
\beq
\label{Max1}
\delta^* V_1 = -\frac{i}{\sqrt2}\Bigl[ \eta D + \ov{\eta D} \Bigr] V_2 \, , \qquad\qquad
\delta^* V_2 = \sqrt2 i \Bigr[ \eta D + \ov{\eta D} \Bigr] V_1
\eeq
provide a representation of $N=2$ supersymmetry with $16_B+16_F$ fields. We may reduce
the supermultiplet by imposing on $V_1$ and $V_2$ constraints consistent with the second
supersymmetry variations: for instance, the single-tensor multiplet is obtained by requiring 
$V_1=L$ and $V_2=\Phi+\ov\Phi$. Another option is to impose a gauge invariance: we may impose that
the theory is invariant under\footnote{For clarity, we use the following convention for field variations:
$\delta^*$ refers to the second ($N=2$) supersymmetry variations of the superfields and component
fields; $\delta_{U(1)}$ indicates
the Maxwell gauge variations; $\delta$ appears for gauge variations of superfields or field 
components related (by supersymmetry) to $\delta b_{\mu\nu} = 2\,\partial_{[\mu} \Lambda_{\nu]}$.}
\beq
\label{Max2}
\delta_{U(1)}\,V_1 = \Lambda_\ell \,, \qquad\qquad 
\delta_{U(1)}\,V_2 = \Lambda_c + \ov\Lambda_c \,,
\eeq
where $\Lambda_\ell$ and $\Lambda_c$ form a single-tensor multiplet,
\beq
\label{Max3}
\Lambda_\ell = \ov\Lambda_\ell \,, \qquad\qquad DD\Lambda_\ell=0,  \qquad\qquad
\ov D_\dalpha \Lambda_c=0,
\eeq
with transformations (\ref{ST4}). Defining the gauge invariant superfields\footnote{Remember that
with this (standard) convention, $\ov W_\dalpha$ is {\it minus} the complex conjugate of $W_\alpha$.}
\beq
\label{Max4}
\begin{array}{rclrcl}
W_\alpha &=& -\frac{1}{4}\,\ov{DD}D_\alpha \, V_2 \,, \qquad&\qquad
\ov W_\dalpha &=& -\frac{1}{4}\, DD \ov D_\dalpha \, V_2 \,,
\crbig
X &=& {1\over2}\, \ov{DD}\, V_1\,, & \ov X &=& {1\over2}\, DD\, V_1,
\end{array}
\eeq
the variations (\ref{Max1}) imply\footnote{There is a phase choice in the definition of $X$: a
phase rotation of $X$ can be absorbed in a phase choice of $\eta$.}
\beq
\label{Max5}
\begin{array}{l}
\delta^*X = \sqrt 2 \, i \, \eta^\alpha W_\alpha ,
\qquad\qquad\qquad\qquad
\delta^*\ov X = \sqrt 2 \, i \, \ov\eta_\dalpha \ov W^\dalpha ,
\crbig
\delta^* W_\alpha =  \sqrt 2 \, i \left[ \frac{1}{4}\eta_\alpha \ov{DD}\,\ov X 
+ i (\sigma^\mu\ov\eta)_\alpha \, \partial_\mu X \right] ,
\crbig
\delta^* \ov W_\dalpha = \sqrt 2 \, i \, \left[ \frac{1}{4} \ov\eta_\dalpha {DD}\, X 
- i (\eta\sigma^\mu)_\dalpha \, \partial_\mu \ov X \right].
\end{array}
\eeq
While $(V_1,V_2)$ describes the $N=2$ supersymmetric extension of the gauge potential 
$A_\mu$, $(W_\alpha, X)$ is the multiplet of the gauge curvature $F_{\mu\nu} =
2\,\partial_{[\mu} A_{\nu]}$ \cite{Maxmult}.

The $N=2$ gauge-invariant Lagrangian depends on the derivatives of a holomorphic prepotential 
${\cal F}(X)$:
\beq
\label{Max6}
\begin{array}{rcl}
{\cal L}_{Max.} &=& {1\over4}\Fint \Bigl[ {\cal F}^{\prime\prime}(X)WW - {1\over2}{\cal F}^\prime(X)
\ov{DD}\,\ov X \Bigr] + {\rm c.c.}
\crbig
&=& {1\over4}\Fint {\cal F}^{\prime\prime}(X)WW + {\rm c.c.}
+ {1\over2} \Dint \Bigl[ {\cal F}^\prime(X)\ov X + \ov{\cal F}^\prime(\ov X) X \Bigr] 
+\partial_\mu(\ldots).
\end{array}
\eeq

In the construction of the Maxwell-multiplet in terms of $X$ and $W_\alpha$, one expects a triplet of 
Fayet-Iliopoulos terms,
\beq
\label{Max7}
{\cal L}_{F.I.}  = - {1\over4} (\xi_1+ ia)\Fint X - {1\over4} (\xi_1-ia)\Fbarint \ov X + \xi_2 \Dint  V_2 ,
\eeq
with real parameters $\xi_1$, $\xi_2$ and $a$. They may generate background values of the auxiliary
components $f_X$ and $d_2$ of $X$ and $V_2$ which in general break both supersymmetries:
\beq
\label{Max8}
\delta^* X= \sqrt2i \, \eta\theta\, \langle d_2 \rangle + \ldots,
\qquad\qquad
\delta^*W_\alpha =\sqrt2i \, \eta_\alpha\, \langle \ov f_X \rangle + \ldots
\eeq
In terms of $V_1$ and $V_2$ however, the relation $X= {1\over2}\ov{DD} V_1$ implies
that $\Im f_X$ is the curl of a three-index antisymmetric tensor (see Subsection \ref{secV1XY}) and
that its expectation value is turned into an integration constant of the tensor field equation 
\cite{ANT, ATT}. As a consequence, 
$$
- {1\over4} (\xi_1+ia)\Fint X - {1\over4} (\xi_1-ia)\Fbarint \ov X = \xi_1 \Dint V_1 + {\rm derivative}
$$
and the Fayet-Iliopoulos Lagrangian becomes
\beq
\label{Max9}
{\cal L}_{F.I.}  = \Dint [\xi_1 V_1 + \xi_2 V_2],
\eeq
with two real parameters only. 

The Maxwell multiplet with superfields $(X,W_\alpha)$ and the single-tensor multiplet
$(Y,\chi_\alpha,\Phi)$ have a simple interpretation in terms of chiral superfields on $N=2$ 
superspace. We will use this formalism to construct their interacting Lagrangians in Section
\ref{secchiralN=2}.

%*****************************************************
\subsection{The Chern-Simons interaction} \label{secCS}
%*****************************************************

With a Maxwell field $F_{\mu\nu} = 2\,\partial_{[\mu} A_{\nu]}$ (in $W_\alpha$) and an antisymmetric 
tensor $b_{\mu\nu}$ (in $\chi_\alpha$ or $L$), one may expect the presence of a 
$b\wedge F$ interaction
$$
\epsilon^{\mu\nu\rho\sigma} b_{\mu\nu}F_{\rho\sigma} =
2\,\epsilon^{\mu\nu\rho\sigma} A_\mu\partial_{\nu}b_{\rho\sigma} + {\rm derivative}.
$$
This equality suggests that its $N=2$ supersymmetric extension also exists in two forms: 
either as an integral over chiral superspace of an expression depending on $\chi_\alpha$,
$W_\alpha$, $X$, $\Phi$ and $Y$, or as a real expression using $L$, $\Phi+\ov\Phi$, 
$V_1$ and $V_2$.

In the `real' formulation, the $N=2$ Chern-Simons term is\footnote{The dimensions in mass unit of our superfields are as follows: $V_1,V_2 : 0$~, $X, Y:1$~, $W_\alpha, \chi_\alpha: 3/2$~, $\Phi,L:2$. The coupling constant $g$ is then dimensionless. }
\beq
\label{CS1}
{\cal L}_{CS} = -g \Dint \Bigl[ LV_2 + (\Phi+\ov\Phi) V_1 \Bigr],
\eeq
with a real coupling constant $g$.
It is invariant (up to a derivative) under the gauge transformations (\ref{Max2}) of $V_1$ and $V_2$
with $L$ and $\Phi$ left inert. Notice that the introduction of Fayet-Iliopoulos terms for $V_1$ and 
$V_2$ corresponds respectively to the shifts $\Phi+\ov\Phi\rightarrow\Phi+\ov\Phi - \xi_1/g$ and
$L\rightarrow L - \xi_2/g$ in the Chern-Simons term.  

The `chiral' version uses the spinorial prepotential $\chi_\alpha$ instead of $L$.
Turning expression (\ref{CS1}) into a chiral integral and using $X={1\over2}\ov{DD}\,V_1$ leads 
to 
\beq
\label{CS3}
{\cal L}_{CS,\,\chi} = g\Fint  \Bigl[ \chi^\alpha W_\alpha + {1\over2} \Phi X  \Bigr]
+ g\Fbarint \Bigl[ -\ov\chi_\dalpha\ov W^\dalpha + {1\over2} \ov\Phi \ov X  \Bigr] ,
\eeq
which differs from ${\cal L}_{CS}$ by a derivative. The chiral version of the Chern-Simons 
term ${\cal L}_{CS,\chi}$ transforms as a derivative under the gauge variation (\ref{ST3}) of 
$\chi_\alpha$. Its invariance under constant shift symmetry of $\Im\Phi$ follows from 
$X={1\over2}\ov{DD}\,V_1$. It does not depend on $Y$.

The consistent Lagrangian for the Maxwell--single-tensor system with Chern-Simons interaction
is then
\beq
\label{CS4}
{\cal L}_{ST} + {\cal L}_{Max.} + {\cal L}_{CS}
\qquad\qquad{\rm or}\qquad\qquad
{\cal L}_{ST} + {\cal L}_{Max.} + {\cal L}_{CS,\,\chi}.
\eeq
The first two contributions include the kinetic terms and self-interactions of the multiplets
while the third describes how they interact. Each of the three terms is separately $N=2$ supersymmetric.

Using a $N=1$ duality, a linear multiplet can be transformed 
into a chiral superfield with constant shift symmetry and the opposite transformation of course exists. Hence, performing both transformations, a single-tensor multiplet Lagrangian $(L,\Phi)$ with constant shift symmetry of the chiral $\Phi$ has a `double-dual' second version. Suppose that
we start with a Lagrangian where Maxwell gauge symmetry acts as a St\"uckelberg gauging
of the single-tensor multiplet:\footnote{Strictly speaking, the coupling constant $g$ in this theory
has dimension (energy)$^2$. There is an irrelevant energy scale involved in the duality transformation
of a dimension two $L$ into a dimension two chiral superfield. Hence, $g$ in Eq.~(\ref{B4}) is again
dimensionless. }
\beq
\label{B1}
{\cal L} = \Dint {\cal H}(L-gV_1, \Phi+\ov\Phi - gV_2) .
\eeq
The shift symmetry of $\Im\Phi$ has been gauged and ${\cal L}$ 
is invariant under gauge transformations (\ref{Max2}) combined with
\beq
\label{B3}
\delta_{U(1)} L = g\Lambda_\ell \,, \qquad\qquad
\delta_{U(1)} \Phi = g\Lambda_c \,,
\eeq
and under $N=2$ supersymmetry if ${\cal H}$ verifies Laplace equation (\ref{ST7}). 
If we perform a double dualization $(L,\Phi+\ov\Phi)\rightarrow (\tilde\Phi + \ov{\tilde\Phi},
\tilde L)$, we obtain the dual theory 
\beq
\label{B4}
\tilde{\cal L} = \Dint \tilde{\cal H}(\tilde L, \tilde\Phi+\ov{\tilde\Phi}) 
+g \Fint \left[ \tilde\chi^\alpha W_\alpha + {1\over2}\tilde\Phi X \right] + {\rm c.c.} ,
\eeq
where $\tilde{\cal H}$ is the result of the double Legendre transformation
\beq
\label{B5}
\tilde{\cal H} (\tilde y, \tilde x) = {\cal H}(x,y) - \tilde xx - \tilde yy.
\eeq
The dual theory is then the sum of the ungauged Lagrangian (\ref{ST7}) and of the Chern-Simons 
coupling (\ref{CS1}). This {\it single-tensor -- single-tensor} duality is actually $N=2$ covariant: if 
${\cal H}$ solves Laplace equation, so does $\tilde{\cal H}$, and every intermediate step of
the duality transformation can be formulated with explicit $N=2$ off-shell supersymmetry.

We have then found two classes of couplings of Maxwell theory to the single-tensor multiplet.
Firstly, using the supersymmetric extension of the $b\wedge F$ coupling, as in Eqs.~(\ref{CS4}).
Secondly, using a St\"uckelberg gauging (\ref{B1}) of the single-tensor kinetic terms. The first
version only is directly appropriate to perform an electric-magnetic duality transformation. However, 
since the second version can always be turned into the first one by a 
single-tensor -- single-tensor duality, electric-magnetic duality of the second version requires this 
preliminary step: both theories have the same `magnetic' dual. 

%***********************************************************
\subsection{The significance of \boldmath{$V_1$}, \boldmath{$X$} and \boldmath{$Y$}} 
\label{secV1XY}
%***********************************************************

In the description of the $N=2$ Maxwell multiplet in terms of two $N=1$ real superfields, 
$V_2$ describes as usual the gauge potential $A_\mu$, a gaugino 
$\lambda_\alpha$ and a real auxiliary field $d_2$ (in Wess-Zumino gauge). We wish to clarify
the significance and the field content of the superfields $V_1$ and $X={1\over2} \ov{DD}V_1$, as 
well as the related content of the chiral superfield $Y$ used in the description in 
terms of the spinorial potential $\chi_\alpha$ of the single-tensor multiplet $(Y,\chi_\alpha,\Phi)$. 

The vector superfield $V_1$ has the $N=2$ Maxwell gauge variation 
$\delta_{U(1)} V_1 = \Lambda_\ell$, with
a real linear parameter superfield $\Lambda_\ell$. In analogy with the Wess-Zumino gauge 
commonly applied to $V_2$, there exists then a gauge where 
\beq
\label{C1}
V_1(x,\theta,\ov\theta) = \theta\sigma^\mu\ov\theta\, v_{1\mu} -{1\over2} \theta\theta \, \ov x -{1\over2} \ov{\theta\theta} \, x
-{1\over\sqrt2} \theta\theta\ov{\theta\psi}_X -{1\over\sqrt2} \ov{\theta\theta}\theta\psi_X 
+{1\over2} \theta\theta\ov{\theta\theta} \, d_1.
\eeq
This gauge leaves a residual invariance acting on the vector field $v_{1\mu}$ only:
\beq
\label{C1a}
\delta_{U(1)} v_1^\mu=\frac{1}{2}\epsilon^{\mu\nu\rho\sigma}\partial_\nu \Lambda_{\rho\sigma} \, .
\eeq
This indicates that the vector $v_1^\mu$ is actually a three-index antisymmetric tensor,
\beq
\label{C2}
v_1^\mu = {1\over6}\epsilon^{\mu\nu\rho\sigma} A_{\nu\rho\sigma} ,
\eeq
with Maxwell gauge invariance 
\beq
\label{C3}
\delta_{U(1)} A_{\mu\nu\rho} = 3\,\partial_{[\mu}\Lambda_{\nu\rho]}.
\eeq
By construction, $X = {1\over2}\ov{DD}V_1$ is gauge invariant. In chiral variables,
\beq
\label{C4}
X(y,\theta) = x + \sqrt2 \, \theta\psi_X - \theta\theta (d_1+i\partial_\mu v_1^\mu).
\eeq
Hence, while $\Re f_X = d_1$,
\beq
\label{C5}
\Im f_X = \partial_\mu v_1^\mu = {1\over24}\epsilon^{\mu\nu\rho\sigma} F_{\mu\nu\rho\sigma},
\qquad\qquad
F_{\mu\nu\rho\sigma} = 4\, \partial_{[\mu}A_{\nu\rho\sigma]}
\eeq
is the gauge-invariant curl of $A_{\mu\nu\rho}$. It follows that the field content (in Wess-Zumino gauge)
of $V_1$ is the second gaugino $\psi_X$, the complex scalar of the Maxwell multiplet $x$, a real
auxiliary field $d_1$ and the three-form field $A_{\mu\nu\rho}$, which corresponds to a single, 
non-propagating component field. The gauge-invariant chiral $X$ includes the four-form curvature 
$F_{\mu\nu\rho\sigma}$.

At the Lagrangian level, the implication of relations  (\ref{C5}) is as follows. Suppose that we compare
two theories with the same Lagrangian ${\cal L}(u)$ but either with $u=\phi$, a real scalar, or with
$u=\partial_\mu V^\mu$, as in Eq.~(\ref{C5}). Since ${\cal L}(\phi)$ does not depend on $\partial_\mu\phi$, the scalar
$\phi$ is auxiliary. The field equations for both theories are
$$
{\partial\over\partial\phi}{\cal L}(\phi) =0 , \qquad\qquad
\partial_\nu \left. {\partial\over\partial u}{\cal L}(u) \right|_{u= \partial_\mu V^\mu} = 0
$$
The second case allows a supplementary integration constant $k$ related to the possible addition of
a `topological' term proportional to $\partial_\mu V^\mu$ to the Lagrangian \cite{ANT, ATT}:
$$
\left. {\partial\over\partial u}{\cal L}(u) \right|_{u= \partial_\mu V^\mu} = k.
$$
In the first case, the same integration constant appears if one considers the following modified 
theory and field equation:
$$
{\cal L}(\phi) - k\,\phi  \qquad\longrightarrow\qquad {\partial\over\partial\phi}{\cal L}(\phi) = k.
$$
Returning to our super-Maxwell case, the relation is $\phi=\Im f_X$ and the modification of the 
Lagrangian is then
\beq
\label{C6}
-k \Im f_X = - {ik\over2} \Fint X + {\rm c.c.}
\eeq
This is the third Fayet-Iliopoulos term, which becomes a `hidden parameter' \cite{ANT} when
using $V_1$ instead of $X$. 

Consider finally the single-tensor multiplet $(Y, \chi_\alpha, \Phi)$ and the supersymmetric extension
of the antisymmetric-tensor gauge symmetry, as given in Eqs.~(\ref{ST3}) and (\ref{ST6}):
$$
\delta Y = -{1\over2}\ov{DD} \Delta^\prime, \qquad\qquad
\delta \chi_\alpha =  {i\over4}\ov{DD} D_\alpha \Delta, \qquad\qquad
\delta \Phi =0.
$$
Using expansion (\ref{C4}), there is a gauge in which $Y$ reduces simply to
\beq
\label{C7}
Y = -i \, \theta\theta\, \Im f_Y
\eeq
and one should identify $\Im f_Y$ as a four-index antisymmetric tensor field, 
\beq
\label{C8}
\Im f_Y = {1\over24} \, \epsilon^{\mu\nu\rho\sigma}C_{\mu\nu\rho\sigma},
\eeq
with residual gauge invariance
\beq
\label{C9}
\delta \, C_{\mu\nu\rho\sigma} = 4\, \partial_{[\mu}\Lambda_{\nu\rho\sigma]}.
\eeq
The antisymmetric tensor $C_{\mu\nu\rho\sigma}$ describes a single field component 
which can be gauged away using
$\Lambda_{\nu\rho\sigma}$. Applying this extended Wess-Zumino 
gauge to the $N=2$ multiplet $(Y,\chi_\alpha,\Phi)$, the fields described by these $N=1$
superfields are as given in the following table. 

\begin{center}
\begin{tabular}{|c|c|c|c|}
\hline
$N=1$ superfield & Field & Gauge invariance & Number of fields \\
\hline 
$\chi_\alpha$ & $b_{\mu\nu}$ & $\delta b_{\mu\nu} = 2\,\partial_{[\mu}\Lambda_{\nu]}$
& $6_B-3_B=3_B$
\\ 
& $C$ && $1_B$
\\
& $\chi_\alpha$ && $4_F$
\\
$\Phi$ & $\Phi$ && $2_B$
\\
& $f_\Phi$ && $2_B$ (auxiliary)
\\
& $\psi_\Phi$ && $4_F$
\\
$Y$ & $C_{\mu\nu\rho\sigma}$ & 
$\delta \, C_{\mu\nu\rho\sigma} = 4\, \partial_{[\mu}\Lambda_{\nu\rho\sigma]}$
& $1_B-1_B=0_B$
\\
\hline
\end{tabular}
\end{center}
The propagating bosonic fields $b_{\mu\nu}$, $C$ and $\Phi$ (four bosonic degrees of freedom) 
have kinetic terms defined by Lagrangian ${\cal L}_{ST}$, Eq.~(\ref{ST7}).

%******************************************************************
\section{Chiral $\bf{ N=2}$ superspace} \label{secchiralN=2}
\setcounter{equation}{0}
%*******************************************************************

Many results of the previous section can be reformulated in terms of chiral superfields on 
$N=2$ superspace. We now turn to a discussion of this framework, including an explicitly 
$N=2$ covariant formulation of electric-magnetic duality.

%*******************************************
\subsection{Chiral $\bf{ N=2}$ superfields}\label{subchiralN=2}
%*******************************************

A chiral superfield on $N=2$ superspace can be written as a function of $y^\mu,
\theta, \tilde\theta$:
\beq
\label{chiral1}
\ov D_\dalpha \,{\cal Z} = \ov {\widetilde D}_\dalpha \,{\cal Z} = 0 \qquad\longrightarrow\qquad
{\cal Z} =  {\cal Z}(y,\theta,\tilde\theta)
\eeq
with $y^\mu = x^\mu - i\theta\sigma^\mu\ov\theta - i\tilde\theta\sigma^\mu\ov{\tilde\theta}$ and 
$\ov D_\dalpha \,y^\mu = \ov {\widetilde D}_\dalpha \,y^\mu = 0$.
Its second supersymmetry variations are
\beq
\label{chiral2}
\delta^*{\cal Z} = i(\eta\tilde Q + \ov \eta\ov{\tilde Q}){\cal Z},
\eeq
with supercharge differential operators $\tilde Q_\alpha$ and $\ov{\tilde Q}_\dalpha$
which we do not need to explicitly write.
It includes four $N=1$ chiral superfields and $16_B+16_F$ component fields and we may use
the expansions
\beq
\label{chiral2b}
\begin{array}{rcl}
{\cal Z}(y,\theta,\tilde\theta) &=& Z(y,\theta) + \sqrt2 \, \tilde\theta^\alpha\omega_\alpha(y,\theta)
 -\tilde\theta\tilde\theta F(y,\theta)
\crbig
&=& Z(y,\theta) + \sqrt2 \, \tilde\theta^\alpha\omega_\alpha(y,\theta)
 -\tilde\theta\tilde\theta \left[ {i\over2}\Phi_{\cal Z}(y,\theta) + {1\over4}\ov{DD}\, \ov Z(y,\theta) \right],
\end{array}
\eeq
where $\tilde\theta$ and $\widetilde D_\alpha$ are the Grassmann coordinates and
the super-derivatives associated with the second supersymmetry. The second supersymmetry 
variations (\ref{chiral2}) are easily obtained by analogy with 
the $N=1$ chiral supermultiplet:
\beq
\label{chiral2c}
\begin{array}{rcl}
\delta^* Z &=& \sqrt2 \, \eta\omega , 
\crbig
\delta^* \omega_\alpha &=& -\sqrt2 [ F \eta_\alpha
+ i (\sigma^\mu\ov\eta)_\alpha \, \partial_\mu Z ]
\,\,=\,\,
-{i\over\sqrt2} \Phi_{\cal Z}\,\eta_\alpha - {\sqrt2\over4} \eta_\alpha \, \ov{DD}\, \ov Z
-\sqrt2 i (\sigma^\mu\ov\eta)_\alpha \partial_\mu Z ,
\crbig
\delta^* F &=& -\sqrt2i\,\partial_\mu\omega\sigma^\mu\ov\eta,
\crbig
\delta^*\Phi_{\cal Z} &=& 2\sqrt2 i \left[\frac{1}{4}\,\ov{DD\eta\omega} 
+ i \partial_\mu\omega\sigma^\mu\ov\eta \right] .
\end{array}
\eeq
We immediately observe that the second expansion (\ref{chiral2b}) leads to the second 
supersymmetry variations (\ref{ST5}) of a single-tensor multiplet $(Y=Z, \chi=\omega, 
\Phi=\Phi_{\cal Z})$. Similarly, the expansion
\beq
\label{chiral4}
{\cal  W}(y, \theta, \tilde \theta) = X(y,\theta) + \sqrt2 i \,\tilde\theta W(y,\theta) 
- \tilde\theta\tilde\theta\, {1\over4}\ov{DD} \ov X (y,\theta),
\eeq
which is obtained by imposing $\Phi_{\cal Z}=0$ in expansion (\ref{chiral2b}), leads to the 
Maxwell supermultiplet (\ref{Max5}) \cite{GSW}. 
The Bianchi identity $D^\alpha W_\alpha = \ov D_\dalpha \ov W^\dalpha$ is required by 
$\delta^*\Phi_{\cal Z}=0$.
The $N=2$ Maxwell Lagrangian (\ref{Max6}) rewrites then as an integral over chiral $N=2$ superspace,
\beq
\label{chiral5}
{\cal L}_{Max.} = {1\over2}\Fint \int d^2\tilde\theta\, {\cal F}({\cal W})
+ {\rm c. c.},
\eeq
and the Fayet-Iliopoulos terms (\ref{Max9}) can be written \cite{IZ}
\beq
\label{DBI12}
{\cal L}_{F.I.} = \Dint [ \xi_1 V_1 + \xi_2V_2]
= -{1\over4}\Fint\int d^2\tilde\theta \left[ \tilde\theta\tilde\theta\,\xi_1 -\sqrt2 i \,\theta\tilde\theta \, \xi_2 
\right] {\cal W} + {\rm c.c.}
\eeq 

Considering the unconstrained chiral superfield (\ref{chiral2b}) with $16_B+16_F$ fields, the reduction
to the $8_B+8_F$ components of the single-tensor multiplet is done by imposing gauge invariance
(\ref{ST3}) and (\ref{ST6}). In terms of $N=2$ chiral superfields, this gauge symmetry is simply
\beq
\label{chiral6}
\delta {\cal Y} =  - \widehat {\cal W}, 
\eeq
where $\widehat {\cal W}$ is a Maxwell $N=2$ superfield parameter (\ref{chiral4}). In terms of $N=1$ superfields,
this is 
\beq
\label{chiral7}
\delta Y =  - \widehat X, \qquad\qquad 
\delta \chi_\alpha =  - i \widehat W_\alpha, \qquad\qquad 
\delta \Phi =  0,
\eeq
as in Eqs.~(\ref{ST3}) and (\ref{ST6}). Hence, a single-tensor superfield ${\cal Y}$ is a chiral
superfield ${\cal Z}$ with the second expansion (\ref{chiral2b}) and with gauge symmetry (\ref{chiral6}).

The chiral version of the Chern-Simons interaction (\ref{CS3}) can be easily written on 
$N=2$ superspace. Using ${\cal Y}$ with gauge invariance (\ref{chiral6}) and ${\cal W}$ to 
respectively describe the single-tensor and the Maxwell multiplets. Then 
\beq
\label{chiral8}
{\cal L}_{CS,\chi} = ig \Fint \int d^2\tilde\theta\, {\cal Y}{\cal W} + {\rm c.c.}
\eeq
It is gauge-invariant since for any pair of Maxwell superfields
\beq
\label{chiral9}
i\Fint \int d^2\tilde\theta\, {\cal W}\widehat{\cal W} + {\rm c.c.}= {\rm derivative}.
\eeq
Notice that the lowest component superfield $Y$ of ${\cal Y}$ does not contribute
to the field equations derived from ${\cal L}_{CS,\chi}$: it only contributes to this Lagrangian
with a derivative.

Finally, a second method to obtain an interactive Lagrangian for the Maxwell--single-tensor 
system is then obvious. Firstly, a generic $N=2$ chiral superfield ${\cal Z}$ can always be
written as
\beq
\label{chiral10}
{\cal Z} = {\cal W} + 2g {\cal Y}.
\eeq
It is invariant under the single-tensor gauge variation (\ref{chiral6}) if one also postulates that
\beq
\label{chiral11}
\delta {\cal W} =  2g\,\widehat {\cal W},
\eeq
which amounts to a $N=2$ St\"uckelberg gauging of the symmetry of the antisymmetric tensor.
With this decomposition, $F_{\mu\nu}$ and $b_{\mu\nu}$ only appear in the $\theta_\alpha
\tilde\theta_\beta$ component of ${\cal Z}$ through the gauge-invariant combination
$F_{\mu\nu}-gb_{\mu\nu}$.
The chiral integral
\beq
\label{chiral12}
{\cal L} = {1\over2}\Fint \int d^2\tilde\theta\, {\cal F}({\cal W} + 2g {\cal Y})
+ {\rm c. c.} + {\cal L}_{ST}
\eeq
provides a $N=2$ invariant Lagrangian describing $16_B+16_F$ (off-shell) interacting fields. 
There exists a gauge in which ${\cal W}=0$, in which case theory (\ref{chiral12}) describes a massive chiral $N=2$ superfield. 

Theory (\ref{chiral12}) is actually related to the Chern-Simons Lagrangian (\ref{CS4}) by electric-magnetic duality, as will be shown below.

%*************************************
\subsection{Electric-magnetic duality}
\label{subsectEMdual}
%*************************************

The description in chiral $N=2$ superspace of the Maxwell multiplet allows 
to derive a $N=2$ covariant version of electric-magnetic duality. The Maxwell Lagrangian (\ref{Max6}) supplemented by the Chern-Simons coupling (\ref{CS3}) can be written
\beq
\label{Belec8}
{\cal L}_{electric} = \Fint\bigint d^2\tilde\theta\, \left[ {1\over2}{\cal F}({\cal W}) 
+ ig{\cal Y}{\cal W}\right] + {\rm c.c.},
\eeq
adding Eqs.~(\ref{chiral5}) and (\ref{chiral8}). Replace then ${\cal W}$ by an unconstrained chiral 
superfield $\hat{\cal Z}$ (with $N=1$ superfields $\hat Z$, $\hat\omega_\alpha$ and $\hat\Phi$) and
introduce a new Maxwell multiplet $\widetilde{\cal W}$ (with 
$N=1$ superfields $\widetilde X$ and $\widetilde W_\alpha$). Using 
$$
\widetilde X = {1\over2} \, \ov{DD}\, \widetilde V_1 \,, \qquad\qquad
\widetilde W_\alpha = -{1\over4} \, \ov{DD}D_\alpha \widetilde V_2 \,,
$$
we have
\beq
\label{Belec9}
\begin{array}{rcl}
i\Fint\bigint d^2\tilde\theta\, \widetilde{\cal W} \hat{\cal Z} + {\rm c.c.}
&=& \Fint \left[ {1\over2}\hat\Phi\widetilde X + \hat\omega\widetilde W\right] + {\rm c.c.}
\crbig
&=& -\Dint\left[ \widetilde V_1(\hat\Phi+\ov{\hat\Phi}) 
+ \widetilde V_2 (D^\alpha\hat\omega_\alpha 
- \ov D_\dalpha\ov{\hat\omega}^\dalpha) \right].
\end{array}
\eeq
Consider now the Lagrangian
\beq
\label{Belec10}
{\cal L} =
\Fint\bigint d^2\tilde\theta\, \left[ {1\over2}{\cal F}(\hat{\cal Z}) 
+ {i\over2} \hat{\cal Z}( \widetilde{\cal W} + 2g{\cal Y})\right] + {\rm c.c.}
\eeq
Invariance under the gauge transformation of the single-tensor superfield,
Eq.~(\ref{chiral6}), requires a compensating gauge variation of 
$\widetilde{\cal W}$, as in Eq.~(\ref{chiral11}).
Eliminating $\widetilde{\cal W}$ leads back to theory (\ref{Belec8}) with 
$\hat{\cal Z}={\cal W}$. This can be seen in two ways. Firstly, the condition
$$
i\Fint\bigint d^2\tilde\theta\, \widetilde{\cal W}\hat{\cal Z} + {\rm c.c.} = {\rm derivative}
$$
leads to $\hat{\cal Z} = {\cal W}$, a $N=2$ Maxwell superfield, up 
to a background value. Secondly, 
using Eqs.~(\ref{Belec9}), we see that $\widetilde V_2$ imposes the Bianchi 
identity on $\hat\omega$ while $\widetilde V_1$ cancels 
$\hat\Phi$ up to an imaginary constant.\footnote{An unconstrained $\widetilde X$ 
would forbid this constant.} We will come back to the (important) role
of a nonzero background value in the next section. For the moment we 
disregard it.

On the other hand, we may prefer to eliminate $\hat{\cal Z}$, 
using its field equation
\beq
\label{Belec11}
{\cal F}^\prime(\hat{\cal Z}) = - i{\cal V} \,,
\qquad\qquad 
{\cal V} \equiv  \widetilde{\cal W} + 2g{\cal Y} \,,
\eeq
which corresponds to a Legendre transformation exchanging variables 
$\hat{\cal Z}$ and ${\cal V}$. Defining
\beq
\label{Belec12}
\widetilde{\cal F}({\cal V}) = {\cal F}(\hat{\cal Z}) + i{\cal V}\hat{\cal Z},
\eeq
we have
\beq
\label{Belec13}
\widetilde{\cal F}^\prime({\cal V}) = i\hat{\cal Z} \,, \qquad\qquad 
{\cal F}^\prime(\hat{\cal Z}) = -i{\cal V} \,, \qquad\qquad
\widetilde{\cal F}^{\prime\prime}({\cal V}){\cal F}^{\prime\prime}(\hat{\cal Z}) 
= 1.
\eeq
The dual (Legendre-transformed) theory is then
\beq
\label{Belec14}
\widetilde {\cal L}_{magnetic} = {1\over2} \Fint\bigint d^2\tilde\theta\,
\widetilde{\cal F}(\widetilde{\cal W}+2g{\cal Y}) + {\rm c.c.}
\eeq
or, expressed in $N=1$ superspace,\footnote{The free, 
canonically-normalized theory corresponds to ${\cal F}({\cal W}) = {1\over2}{\cal W}^2$
and $\widetilde {\cal F}({\cal V}) = {1\over2}{\cal V}^2$.}
\beq
\label{Belec14b}
\begin{array}{rcl}
\widetilde {\cal L}_{magnetic}
&=& {1\over4} \Fint\Bigl[ \widetilde{\cal F}^{\prime\prime}(\widetilde X+2gY) 
\, (\widetilde W-2ig\chi)^\alpha(\widetilde W-2ig\chi)_\alpha
\crbig
&& \hspace{1.0cm}  -{1\over2}\widetilde{\cal F}^\prime(\widetilde X+2gY) \, \ov{DD}(\ov{\widetilde X}
+2g\ov Y) 
 - 2ig\,\widetilde{\cal F}^\prime(\widetilde X+2gY)\Phi \Bigr] + {\rm c.c.}
\end{array}
\eeq
We then conclude that the presence of the Chern-Simons term in the 
electric theory induces a St\"uckelberg gauging in the dual magnetic 
theory.

As explained in Ref.~\cite{IZ}, the situation changes when Fayet-Iliopoulos terms (\ref{DBI12}) 
are present in the electric theory. In the magnetic theory
coupled to the single-tensor multiplet, with Lagrangian (\ref{Belec14b}),
the gauging $\delta\widetilde{\cal W} = 2g\widehat{\cal W}$ forbids
Fayet-Iliopoulos terms for the magnetic Maxwell superfields $\widetilde 
V_1$ and $\widetilde V_2$. Spontaneous supersymmetry breaking by 
Fayet-Iliopoulos terms in the electric theory finds then a different origin in 
the magnetic dual.

For our needs, we only consider the Fayet-Iliopoulos term induced
by $V_1$, {\it i.e.}~we add 
\beq
\label{Belec15}
{\cal L}_{FI} = \xi_1 \int d^4\theta\, V_1 = -{1\over4} \xi_1 
\Fint\int d^2\tilde\theta\, \tilde\theta\tilde\theta \,{\cal W}
+{\rm c.c.}
\eeq
to ${\cal L}_{electric}$, Eq.~(\ref{Belec8}). In turn, this amounts to
add 
$$
 -{1\over4} \xi_1 \Fint\int d^2\tilde\theta\, \tilde\theta\tilde\theta \, \hat{\cal Z}
+{\rm c.c.}
$$
to theory (\ref{Belec10}). But, in contrast to expression 
(\ref{Belec15}), this
modification is not invariant under the second supersymmetry: according to
the first Eq.~(\ref{chiral2c}), its $\delta^*$ variation
$$
 -{\sqrt2\over4} \xi_1 \Fint \eta\omega
+{\rm c.c.}
$$
is not a derivative.\footnote{It would be a derivative if $\omega_\alpha$
would be replaced by the Maxwell superfield $W_\alpha$, as in 
Eq.~(\ref{Belec15}).} To restore $N=2$ supersymmetry, we must 
deform the $\delta^*$ variation of $\widetilde W_\alpha - 2ig \chi_\alpha$
into
\beq
\label{Belec16}
\delta^*_{deformed}(\widetilde W_\alpha - 2ig \chi_\alpha) =
{1\over\sqrt2}\xi_1 \eta_\alpha +
\delta^*(\widetilde W_\alpha - 2ig \chi_\alpha),
\eeq
the second term being the usual, undeformed, variations 
(\ref{Max5}) and (\ref{ST5}). Hence, the magnetic theory has a goldstino
fermion and linear $N=2$ supersymmetry partially breaks to $N=1$, as a consequence of 
the electric Fayet-Iliopoulos term.
Concretely, the magnetic theory is now
\beq
\label{Belec17}
\begin{array}{rcl}
\widetilde {\cal L}_{magnetic} &=& {1\over2} \Fint\bigint d^2\tilde\theta\,
\widetilde{\cal F}\Bigl(\widetilde{\cal W}+2g{\cal Y} + {i\over2}\xi_1
\tilde\theta\tilde\theta\Bigr) + {\rm c.c.}
\crbig
&=& {1\over2} \Fint\bigint d^2\tilde\theta\, \left[
\widetilde{\cal F}\Bigl(\widetilde{\cal W}+2g{\cal Y} \Bigr) 
+ {i\over2}\xi_1 \tilde\theta\tilde\theta\,
\widetilde{\cal F}^\prime\Bigl(\widetilde{\cal W}+2g{\cal Y} \Bigr) 
\right] + {\rm c.c.}
\crbig
&=& \left[ {1\over2} \Fint \bigint d^2\tilde\theta\, 
\widetilde{\cal F}\Bigl(\widetilde{\cal W}+2g{\cal Y} \Bigr) 
+ {i\over4}\xi_1 \Fint
\widetilde{\cal F}^\prime\Bigl(\widetilde X +2gY \Bigr) \right]
+ {\rm c.c.}
\end{array}
\eeq
One easily checks that $N=2$ supersymmetry holds, using the deformed variations (\ref{Belec16}).

%******************************************************
\section{Nonlinear \boldmath{$N=2$} supersymmetry and the DBI action} \label{secDBI}
\setcounter{equation}{0}
%*******************************************************

In the previous sections, we have developed various aspects of the coupling of a
Maxwell multiplet to a single-tensor multiplet in linear $N=2$ supersymmetry. With these tools,
we can now address our main subject: show how a Dirac-Born-Infeld Lagrangian (DBI) coupled to
the single-tensor multiplet arises from non-linearization of the second supersymmetry.

It has been observed that the DBI Lagrangian with nonlinear second 
supersymmetry can be derived by solving a constraint invariant under $N=2$ supersymmetry 
imposed on the super-Maxwell theory \cite{BG, RT}. We start with a summary of this result, following 
mostly Ro\v cek and Tseytlin \cite{RT}, and we then generalize the method to incorporate the fields of the
single-tensor multiplet. 

%*******************************************************************
\subsection{The \boldmath{$N=2$} super-Maxwell DBI theory}
%*******************************************************************

The constraint imposed on the $N=2$ Maxwell chiral superfield 
${\cal W}$ is \cite{RT}\footnote{See also Ref.~\cite{R} and very recently Ref.~\cite{KS} in the context of 
$N=1$ supersymmetry.}
\beq
\label{DBI5}
{\cal W}^2- {1\over\kappa} \tilde\theta\tilde\theta \, {\cal W}
= \left( {\cal W}  - {1\over2\kappa} \tilde\theta\tilde\theta \right)^2 = 0.
\eeq
It imposes a relation between the super-Maxwell Lagrangian superfield 
${\cal W}^2$ and the Fayet-Iliopoulos `superfield' $\tilde\theta\tilde\theta{\cal W}$, Eq.~(\ref{Belec15}). 
The real scale parameter $\kappa$ has dimension (energy)$^{-2}$.
In terms of $N=1$ superfields, the constraint is equivalent to
\beq
\label{DBI6}
X^2 =0 , \qquad\qquad XW_\alpha =0 , \qquad\qquad
WW - {1\over2} X\ov{DD}\ov X =  {1\over\kappa} X.
\eeq
The third equality leads to
\beq
\label{DBI7}
X = {2 \, WW \over  {2\over\kappa} + \ov{DD}\ov X}
\eeq
which, since $W_\alpha W_\beta W_\gamma =0$, implies the first two conditions. Solving 
the third constraint amounts to express $X$ as a function of $WW$ \cite{BG}\footnote{See Appendix B.}. 
The DBI theory is then obtained using as Lagrangian the Fayet-Iliopoulos
term (\ref{Belec15}) properly normalized:
\beq
\label{DBI8}
{\cal L}_{DBI} = {1\over4\kappa} \Fint X + {\rm c.c}
= {1\over8\kappa^2} \left[ 1- \sqrt{- {\rm det}(\eta_{\mu\nu} + 2\sqrt2\kappa F_{\mu\nu} )} \right] 
+ \ldots
\eeq
The constraints (\ref{DBI5}) and (\ref{DBI6}) are not invariant under 
the second linear supersymmetry, with variations $\delta^*$. However, 
one easily verifies that the three constraints (\ref{DBI6}) are invariant under 
the deformed, nonlinear variation 
\beq
\label{DBI9}
\delta^*_{deformed}  W_\alpha =  \sqrt 2 \, i \left[ {1\over2\kappa}\eta_\alpha
+\frac{1}{4}\eta_\alpha \ov{DD}\,\ov X 
+ i (\sigma^\mu\ov\eta)_\alpha \, \partial_\mu X \right] ,
\eeq
with $\delta^*X$ unchanged. 
The deformation preserves the $N=2$ supersymmetry algebra. It indicates 
that the gaugino spinor in $W_\alpha = -i\lambda_\alpha + \ldots$ 
transforms inhomogeneously,
$\delta^*\lambda_\alpha = -{1\over\sqrt2\kappa}\, \eta_\alpha+\ldots$, 
like a goldstino for the breaking of the second supersymmetry. 
In other words, at the level of the $N=2$ chiral superfield ${\cal W}$,
$$
\delta^*_{deformed} \, {\cal W} = -{1\over\kappa} \tilde\theta\eta + 
i\left(\eta\tilde Q + \ov\eta\ov{\tilde Q}\right){\cal W}
= i \left(\eta\tilde Q + \ov\eta\ov{\tilde Q}\right) \left({\cal W} - {1\over2\kappa} \tilde\theta
\tilde\theta\right).
$$
The deformed second supersymmetry variations $\delta^*_{deformed}$ 
act on ${\cal W}$ as the usual variations $\delta^*$ act on the shifted superfield
${\cal W}  - {1\over2\kappa} \tilde\theta\tilde\theta$. In fact, this superfield
transforms like a chiral $N=2$ superfield (\ref{chiral2b}) with $Z=X$, $\omega_\alpha = iW_\alpha$
verifying the Bianchi identity and with $\Phi_{\cal Z}=-i/\kappa$. The latter background value of 
$\Phi_{\cal Z}$ may be viewed as the source of the partial breaking of linear supersymmetry.

Hence, the scale parameter $\kappa$ introduced in the nonlinear constraint (\ref{DBI5})
appears as the scale parameter of the DBI Lagrangian and also as the order parameter of 
partial supersymmetry breaking. The Fayet-Iliopoulos term (\ref{DBI8}) has in principle an
arbitrary coefficient $-\xi_1/4$, as in Eq.~(\ref{Max9}). We have chosen $\xi_1= -\kappa^{-1}$ to canonically
normalize gauge kinetic terms.

The DBI Lagrangian is invariant under electric-magnetic duality.\footnote{For instance, in the 
context of D3-branes of IIB superstrings, see Ref.~\cite{T}. Our procedure is inspired by 
Ref.~\cite{RT}.} In our $N=2$ case, the invariance
is easily established in the language of $N=2$ superspace. 
We first include the constraint as a field equation of the Lagrangian:
\beq
\label{DBIEM1}
{\cal L}_{DBI} = \Fint\int d^2\tilde\theta \left[ {1\over4\kappa}\tilde\theta\tilde\theta\,{\cal W}
+ {1\over4}\Lambda\left( {\cal W} - {1\over2\kappa}\tilde\theta\tilde\theta \right)^2\,  \right] + {\rm c.c.}
\eeq
The field equation of the $N=2$ superfield $\Lambda$ enforces (\ref{DBI5}). We then introduce
two unconstrained $N=2$ chiral superfields $U$ and $\Upsilon$ and the modified Lagrangian
$$
{\cal L}_{DBI} = \Fint\int d^2\tilde\theta \left[ {1\over4\kappa}\tilde\theta\tilde\theta\,{\cal W}
+ {1\over4}\Lambda U^2 
- {1\over2} \Upsilon\left(U - {\cal W} + {1\over2\kappa}\tilde\theta\tilde\theta\right)
 \right] + {\rm c.c}.
$$
Since the Lagrange multiplier $\Upsilon$ imposes $U={\cal W} - {1\over2\kappa}\tilde\theta\tilde\theta$,
the equivalence with (\ref{DBIEM1}) is manifest. But we may also eliminate ${\cal W}$ which only 
appears linearly in the last version of the theory. The result is 
$$
\Upsilon = -i\widetilde{\cal W} - {1\over2} \left( {1\over\kappa} - i\zeta \right)\tilde\theta\tilde\theta
$$
where $\widetilde{\cal W}$ is a Maxwell $N=2$ superfield dual to ${\cal W}$ and $\zeta$ an 
arbitrary real constant. 
As in Subsection \ref{subsectEMdual}, $N=2$ supersymmetry of the theory with a Fayet-Iliopoulos 
term requires a nonlinear deformation of the $\delta^*$ variation of $\widetilde{\cal W}$:
$\widetilde{\cal W}  - {i\over2} \left( {1\over\kappa} - i\zeta \right)\tilde\theta\tilde\theta$ should be a 
`good' $N=2$ chiral superfield. Replacing $\Upsilon$ in the Lagrangian and taking $\zeta=0$ leads to
$$
{\cal L}_{DBI} = \Fint\int d^2\tilde\theta \left[  {1\over4}\Lambda U^2 
+ {i\over2} U \left[\widetilde{\cal W} - {i\over2\kappa} \tilde\theta\tilde\theta \right]
+ {i\over4\kappa}\widetilde{\cal W} \, \tilde\theta\tilde\theta 
 \right] + {\rm c.c}.
$$
Finally, eliminating $U$ gives the magnetic dual
\beq
\label{DBIEM2}
{\cal L}_{DBI} = \Fint\int d^2\tilde\theta \left[  {1\over4\Lambda} 
\left(\widetilde{\cal W} - {i\over2\kappa} \tilde\theta\tilde\theta \right)^2
+ {i\over4\kappa}\widetilde{\cal W} \, \tilde\theta\tilde\theta 
 \right] + {\rm c.c}.
\eeq
One easily verifies that the resulting theory has the same expression as the initial `electric'  
theory (\ref{DBI8}). The Lagrange multiplier $\Lambda^{-1}$ imposes constraint
(\ref{DBI5}) to $-i\widetilde{\cal W}$, which reduces to Eq.~(\ref{DBI7}) applied to $-i\widetilde X$.
The Lagrangian is then given by the Fayet-Iliopoulos term for this superfield.

%*******************************************
\subsection{Coupling the DBI theory to a single-tensor multiplet:\\ 
a super-Higgs mechanism without gravity}\label{couplingDBIst}
%*******************************************

The $N=2$ super-Maxwell DBI theory is given by a Fayet-Iliopoulos term
for a Maxwell superfield submitted to the quadratic constraint (\ref{DBI5}),
which also provides the source of partial supersymmetry breaking. 
The second supersymmetry is deformed by the constraint: it is 
${\cal W} - {1\over2\kappa}\tilde\theta\tilde\theta$ which
transforms as a regular $N=2$ chiral superfield. Instead of expression
(\ref{chiral8}), we are thus led to consider the following Chern-Simons 
interaction with the single-tensor multiplet:
\beq
\label{DBI}
\begin{array}{rcl}
{\cal L}_{CS, def.} &=& ig\Fint\bigint d^2\tilde\theta \, {\cal Y}\left( {\cal W}
- {1\over2\kappa}\tilde\theta\tilde\theta\right)
+ {\rm c.c.}
\crbig
&=& g\Fint\left[ {1\over2}\Phi X + \chi^\alpha W_\alpha
- {i\over2\kappa} Y \right]
+ {\rm c.c.} + {\rm derivative.}
\end{array}
\eeq
The new term induced by the deformation of $\delta^*W_\alpha$ 
is proportional to the four-form field described by the chiral superfield $Y$, 
as explained in Subsection \ref{secV1XY} [see Eq.~(\ref{C8})].
This modified Chern-Simons interaction, invariant under the deformed second supersymmetry 
variations, may be simply added to the Maxwell DBI theory (\ref{DBIEM1}). 
We then consider the Lagrangian
\beq
\label{DBIa}
{\cal L}_{DBI} = \Fint\int d^2\tilde\theta \left[ ig{\cal Y}
\left( {\cal W} - {1\over2\kappa}\tilde\theta\tilde\theta\right)
-{1\over4} \xi_1\tilde\theta\tilde\theta\,{\cal W}
+ {1\over2} \Lambda\left( {\cal W} - {1\over2\kappa}\tilde\theta\tilde\theta \right)^2\,  \right] + {\rm c.c.},
\eeq
for the constrained Maxwell and single-tensor multiplets, keeping the Fayet-Iliopoulos coefficient
$\xi_1$ arbitrary. For a coherent theory with a propagating single-tensor multiplet,
a kinetic Lagrangian ${\cal L}_{ST}$ [Eq.~(\ref{ST7})] should also be added. 
Since
$$
\Fint\int d^2\tilde\theta \left[ ig{\cal Y}{\cal W}
-{1\over4}  \xi_1\tilde\theta\tilde\theta\,{\cal W}\right] + {\rm c.c.}
= \Fint \left[ g\,\chi W +{g\over2}\Phi X -{1\over4} \xi_1 X \right] 
+ {\rm c.c.} + {\rm deriv.},
$$
we see that the Fayet-Iliopoulos term is equivalent to a constant real 
shift of $\Phi$ which, according to variations (\ref{ST5}), partially breaks supersymmetry. 
We will choose to expand $\Phi$ around $\langle\Phi\rangle=0$ and keep $\xi_1\ne0$.

Again, the constraint (\ref{DBI5}) imposed by the Lagrange multiplier 
$\Lambda$ can be solved to express $X$ as a function of $WW$: $X=X(WW)$.
The result is \cite{BG}
\beq
\label{XWWis}
X(WW) =
\kappa WW - \kappa^3 \ov{DD} \left[ { WW \ov{WW} \over 
1 + \kappa^2A + \sqrt{1 +2\kappa^2A + \kappa^4B^2}} \right],
\eeq
where $A$ and $B$ are defined in Appendix B. The DBI Lagrangian coupled to the single-tensor
multiplet reads then
\beq
\label{DBIb}
{\cal L}_{DBI} = 
\Fint\left[ {1\over4} \left(2g\Phi - \xi_1\right) X(WW) + g\chi^\alpha W_\alpha
- {ig\over2\kappa} Y \right]
+ {\rm c.c.} + {\cal L}_{ST}.
\eeq
The bosonic Lagrangian depends on a single auxiliary field\footnote{
Since $X(WW)|_{\theta=0}$ is a function of fermion bilinears, 
the auxiliary $f_\Phi$ does not contribute to the 
bosonic Lagrangian and $\chi_\alpha$ does not include any auxiliary field.}, $d_2$ in $W_\alpha$ 
or $V_2$:
\beq
\label{DBIc}
\begin{array}{rcl}
{\cal L}_{DBI,\, bos.} &=&  {1\over8\kappa}(2g\Re\Phi - \xi_1)
\left( 1-\sqrt{-8\kappa^2 d_2^2-\det (\eta_{\mu\nu}+2\sqrt 2\kappa\,F_{\mu\nu})}\right)
- {g\over2}Cd_2 
\crbig
&& + g\epsilon^{\mu\nu\rho\sigma}\left( {\kappa\over4}\Im\Phi F_{\mu\nu}F_{\rho\sigma}
- {1\over4}b_{\mu\nu}F_{\rho\sigma} 
+ {1\over24\kappa} C_{\mu\nu\rho\sigma} 
\right) + {\cal L}_{ST, \, bos.}.
\end{array}
\eeq
The real scalar field $C$ is the lowest component of the linear superfield $L$. Contrary to
$\langle\Phi\rangle$, its background value is allowed by $N=2$ supersymmetry. However,
a non-zero $\langle C \rangle$ would induce a non-zero $\langle d_2\rangle$ which would  
spontaneously break the residual $N=1$ linear supersymmetry. This is visible in the bosonic 
action which, after elimination of
\beq
\label{d2elec}
d_{2, \,bos.} = {gC\over2\kappa}  \sqrt{-\det (\eta_{\mu\nu}+2\sqrt 2\kappa\,F_{\mu\nu}) \over 
(2g\Re\Phi - \xi_1)^2 + 2g^2 C^2},
\eeq
becomes
\beq
\label{DBId}
\begin{array}{rcl}
{\cal L}_{DBI,\, bos.} &=&  {1\over8\kappa}(2g\Re\Phi - \xi_1)\left[ 1 -
\sqrt{1 + {2g^2 C^2 \over (2g\Re\Phi - \xi_1)^2}}
\sqrt{-\det (\eta_{\mu\nu}+2\sqrt 2\kappa\,F_{\mu\nu})}
\right]
\crbig
&& + g\epsilon^{\mu\nu\rho\sigma}\left( {\kappa\over4}\Im\Phi F_{\mu\nu}F_{\rho\sigma}
- {1\over4}b_{\mu\nu}F_{\rho\sigma} 
+ {1\over24\kappa} C_{\mu\nu\rho\sigma} 
\right) + {\cal L}_{ST, \, bos.}.
\end{array}
\eeq
First of all, as expected, the theory includes a DBI Lagrangian for the Maxwell field strength 
$F_{\mu\nu}$, with scale $\sim \kappa$. With the Chern-Simons coupling to the single-tensor 
multiplet, the DBI term acquires a field-dependent coefficient,
\beq
- {1\over8\kappa} \sqrt{(2g\Re\Phi - \xi_1)^2 + 2g^2 C^2Ê} \,
\sqrt{-\det (\eta_{\mu\nu}+2\sqrt 2\kappa\,F_{\mu\nu})}.
\eeq
It also includes a $F\wedge F$ term which respects the axionic shift symmetry of $\Im\Phi$, a
$b\wedge F$ coupling induced by (linear) $N=2$ supersymmetry and a `topological' $C_4$ term
induced by the nonlinear deformation. These terms are strongly reminiscent of those found when
coupling a D-brane Lagrangian to IIB supergravity. The contribution of the four-form can be 
eliminated by a gauge choice of the single-tensor symmetry (\ref{C9}). We have however 
insisted on keeping off-shell (deformed) $N=2$ supersymmetry, hence the presence of this term. 

The theory also includes a semi-positive scalar potential\footnote{We only consider 
$2g\Re\Phi - \xi_1 > 0$, in order to have well-defined positive gauge kinetic terms.}
\beq
\label{DBIpot}
V(C,\Re\Phi) = {2g\Re\Phi - \xi_1 \over8\kappa}
\left[ \sqrt{1 + {2g^2 C^2 \over (2g\Re\Phi - \xi_1)^2}} - 1 \right]
\eeq
which vanishes only if $C$ is zero.\footnote{With respect to $\Re\Phi$, the potential is stationary, 
${\partial V\over\partial\Re\Phi}=0$, only if $C=0$. All local minima are then characterized by
$C=0$ and $\Re\Phi$ arbitrary and are then (supersymmetric) global minima.}
The scalar potential determines then $\langle C \rangle=0$ but leaves $\Re\Phi$ arbitrary. Since
$$
\langle d_2 \rangle = {g\langle C\rangle\over2\kappa} \,
\Bigl\langle (2g\Re\Phi - \xi_1)^2 + 2g^2 C^2 \Bigr\rangle^{-1/2},
$$
the vacuum line $\langle C \rangle=0$ is compatible with linear $N=1$ and deformed 
second supersymmetry. While $\Phi$ is clearly massless, $C$ has a mass term
$$
- {1\over2}M_C^2 \, C^2
= - {g^2\over4\kappa(2\Re\Phi - \xi_1)}  C^2.
$$
The same mass is acquired by the $U(1)$ gauge field coupled to the antisymmetric tensor 
$b_{\mu\nu}$, and by the goldstino (the $U(1)$ gaugino in $W_\alpha$) that forms a Dirac spinor 
with the fermion of the linear multiplet $\chi_\alpha$. In other words, the Chern-Simons coupling 
$\chi W$ pairs the Maxwell goldstino with the linear multiplet to form a massive vector, while the 
chiral multiplet $\Phi$ remains massless with no superpotential. 

At $\langle C \rangle=\langle \Re\Phi \rangle=0$, gauge kinetic terms are canonically normalized if
$\xi_1 = -\kappa^{-1}$. The Maxwell DBI theory (\ref{DBI8}) is of course recovered when the 
Chern-Simons interaction decouples with $g=0$.
Notice finally that the kinetic terms ${\cal L}_{ST}$ of the single-tensor multiplet are given by 
Eq.~(\ref{ST7}), as with linear $N=2$ supersymmetry.
Since the nonlinear deformation of the second supersymmetry does not affect $\delta^*L$ or 
$\delta^*\Phi$ even if $\langle\Re\Phi\rangle\ne0$, the
function ${\cal H}$ remains completely arbitrary.

The phenomenon described above provides a first instance of a super-Higgs mechanism without 
gravity: the nonlinear goldstino multiplet is `absorbed' by the linear multiplet to form a massive 
vector $N=1$ superfield. One may wonder how this can happen without gravity; normally one 
expects that the goldstino can be absorbed only by the gravitino in local supersymmetry. The 
reason of this novel mechanism is that the goldstino sits in the same multiplet of the linear supersymmetry as a gauge field which has a Chern-Simons interaction with the tensor multiplet. 
This will become clearer in Section \ref{secQED}, where we will show by a change of variables  
that this coupling is equivalent to an ordinary gauge interaction with a charged hypermultiplet, 
providing non derivative gauge couplings to the goldstino. Actually, this particular super-Higgs
mechanism is an explicit realization of a phenomenon known in string theory where
the $U(1)$ field of the D-brane world-volume becomes 
in general massive due to a Chern-Simons interaction with the R--R antisymmetric tensor of a 
bulk hypermultiplet.\footnote{This can be avoided in the orientifold case: the $N=2$ bulk
supermultiplets are truncated by the orientifold projection.}

We have chosen a description in terms of the single-tensor multiplet because it admits an 
off-shell formulation well adapted to our problem.
Our DBI Lagrangian (\ref{DBIa}), supplemented with kinetic terms ${\cal L}_{ST}$, admits however 
several duality transformations. Firstly, since it only depends on ${\cal W}$, we may perform an
electric-magnetic duality transformation, as described in Subsection \ref{secmagndual}. 
Then, for any choice
of ${\cal L}_{ST}$, we can transform the linear $N=1$ superfield $L$ into a chiral $\Phi^\prime$.
The resulting theory is a hypermultiplet formulation with superfields $(\Phi,\Phi^\prime)$ and 
$N=2$ supersymmetry realized only on-shell. As already
explained in Subsection \ref{secCS}, the $b\wedge F$ interaction is replaced by a St\"uckelberg 
gauging of the axionic shift symmetry of the new chiral $\Phi^\prime$: the K\"ahler potential of 
the hypermultiplet formulation is a function of $\Phi^\prime+ \ov\Phi^\prime - gV_2$. 
Explicit formulae  are given in the next subsection and in 
Section \ref{secQED} we will use this mechanism in the case of nonlinear $N=2$ QED. Finally, 
if kinetic terms ${\cal L}_{ST}$ also respect the shift symmetry of $\Im\Phi$, the chiral $\Phi$ can be
turned into a second linear superfield $L^\prime$, leading to two formulations
which are also briefly described below.

%%%%%%%%%
\subsection{Hypermultiplet, double-tensor and single-tensor \\ dual formulations}  
\label{sechyperform}
%%%%%%%%%%

As already mentioned, using the single-tensor multiplet is justified by the existence of an 
off-shell $N=2$ formulation. The hypermultiplet formulation, with two $N=1$ chiral superfields,
is however more familiar and the first purpose of this subsection is to translate our results into this
formalism.
In the DBI theory (\ref{DBIb}), the linear superfield $L$ only appears in
$$
\begin{array}{l}
{\cal L}_{ST} + g\Fint \chi^\alpha W_\alpha + {\rm c.c.}
= \Dint \left[{\cal H} (L, \Phi, \ov \Phi) + g L V_2\right] +{\rm derivative.}
\end{array}
$$
These contributions are not invariant under $\delta^*$ variations: the nonlinear deformation 
acts on $W_\alpha$ and on $V_2$. Nevertheless,
the linear superfield can be transformed into a new chiral superfield $\Phi^\prime$.
The resulting `hypermultiplet formulation' has Lagrangian
\beq
\label{hyp1}
\begin{array}{rcl}
{\cal L}_{DBI,\, hyper.} &=& \Dint {\cal K}\Bigl(\Phi^\prime+\ov\Phi^\prime - gV_2 , \Phi, \ov\Phi \Bigr)
\crbig
&&+ \Fint\left[ {1\over4}\left(2g\Phi - \xi_1\right) X(WW) 
- {ig\over2\kappa} Y \right]
+ {\rm c.c.}
\end{array}
\eeq
The K\"ahler potential is given by the Legendre transformation
\beq
\label{hyp2}
{\cal K} (\Phi^\prime+\ov\Phi^\prime , \Phi, \ov\Phi )= 
{\cal H} (U,\Phi, \ov\Phi ) - U(\Phi^\prime + \ov \Phi^\prime),
\eeq
where $U$ is the solution of
\beq
\label{hyp3}
{\partial\over\partial U}{\cal H} (U,\Phi, \ov\Phi ) = \Phi^\prime + \ov \Phi^\prime.
\eeq
In the single-tensor formulation, $N=2$ supersymmetry implies that ${\cal H}$ solves Laplace equation.
As a result of the Legendre transformation, the determinant of ${\cal K}$ is constant and the
metric is hyperk\"ahler \cite{LR}. It should be noted that the Legendre transformation defines the
new auxiliary scalar $f_{\Phi^\prime}$ of $\Phi^\prime$ according to
\beq
\label{hyp4}
f_{\Phi^\prime} = \left({\partial^2  {\cal H} \over \partial U\partial \Phi}\right)_{\theta=0} \, f_\Phi.
\eeq
Hence, the hypermultiplet formulation has the same number of independent auxiliary fields as the 
single-tensor version: $d_2$ and $f_\Phi$. 

The second supersymmetry variation $\delta^*$ of $\Phi^\prime$ is also defined by transformation (\ref{hyp3}): in the hypermultiplet formulation, $N=2$ is realized on-shell only, using the Lagrangian function. The nonlinear deformation of variations $\delta^*$ acts on $V_2$. Since 
$W_\alpha = -{1\over4}\ov{DD}D_\alpha V_2$, Eq.~(\ref{DBI9}) indicates that
$$
\delta^* V_2 = {i\over\sqrt2\kappa}(\ov{\theta\theta}\theta\eta - \theta\theta\ov{\theta\eta})
+ \sqrt2i \, (\eta D + \ov{\eta D}) V_1.
$$ 
The $\kappa$-dependent term in the $\delta^*$ variation of the K\"ahler potential term in
${\cal L}_{DBI,\, hyper.}$ is then the same as the $\kappa$-dependent part in 
$g\,\delta^*\int d^2\theta\,\chi^\alpha W_\alpha + {\rm c.c}$, which is compensated by the 
variation of the four-form field. This can again be shown using 
Eqs.~(\ref{hyp2}) and (\ref{hyp3}).
This hypermultiplet formulation will be used in Section \ref{secQED}, on the example of 
nonlinear DBI QED with a charged hypermultiplet.

For completeness, let us briefly mention two further formulations of the same DBI theory, using either 
a double-tensor, or a dual single-tensor $N=2$ multiplet. These possibilities appear if Lagrangian
(\ref{DBIb}) has a second shift symmetry of $\Im\Phi$. This is the case if the single-tensor kinetic
Lagrangian has this isometry:
$$
{\cal L}_{ST} = \Dint {\cal H} (L, \Phi+\ov \Phi) .
$$
We may then transform $\Phi$
into a linear superfield $L^\prime$ using a $N=1$ duality transformation. 
Keeping $L$ and turning $\Phi$ into $L^\prime$ leads to a double-tensor
formulation with superfields $(L,L^\prime)$. The Lagrangian has the form
\beq
\label{DT}
{\cal L}_{DT} = \Dint {\cal G}\Bigl(L,L^\prime - g V_1(WW) \Bigr) 
- \Fint\left[ {1\over4} \xi_1 X(WW) - g\chi^\alpha W_\alpha
+ {ig\over2\kappa} Y \right]
+ {\rm c.c.} 
\eeq
The function ${\cal G}$ is the Legendre transform of ${\cal H}$ with respect to its second variable 
$\Phi+\ov\Phi$ and the real superfield $V_1(WW)$ is defined by the equation
\beq
\label{V1WWis}
X(WW) = {1\over2}\ov{DD}\, V_1(WW).
\eeq
It includes the DBI gauge kinetic term in its $d_1$ component and the Lagrangian depends on the
new tensor $b_{\mu\nu}^\prime$ through the combination $3\,\partial_{[\mu} b^\prime_{\nu\rho]}
- g\,\omega_{\mu\nu\rho}$, where $\omega_{\mu\nu\rho}= 3\,A_{[\mu}F_{\nu\rho]}$ is the 
Maxwell Chern-Simons form. 

Finally, turning $\Phi$ and $L$ into $L^\prime$ and $\Phi^\prime$, leads to another single-tensor theory
with a St\"uckelberg gauging of both $\Phi^\prime$ and $L^\prime$, as in theory (\ref{B1}). In this case,
the Lagrangian is
\beq
\label{STprime}
{\cal L}_{ST^\prime} = \Dint \widetilde{\cal H} \Bigl(\Phi^\prime+\ov\Phi^\prime - gV_2, 
L^\prime - g V_1(WW) \Bigr)
- \Fint\left[ {1\over4}\xi_1 X(WW) 
+ {ig\over2\kappa} Y \right]
+ {\rm c.c.} 
\eeq
While in the double-tensor theory (\ref{DT}) the second nonlinear supersymmetry only holds
on-shell, it is valid off-shell in theory (\ref{STprime}).
The function $\widetilde{\cal H}$ verifies Laplace equation, as required by $N=2$ linear 
supersymmetry.\footnote{See Eq.~(\ref{ST7}).} Using the supersymmetric Legendre transformation, 
one can show that the nonlinear deformation of $\delta^*V_2$, which affects $\delta^*\widetilde{\cal H}$,
is again balanced by the variation of the four-form superfield $Y$.

%%%%%%%%%
\subsection{The magnetic dual}  \label{secmagndual}
%%%%%%%%%

To perform electric-magnetic duality on theory (\ref{DBIa}), we first replace it with 
\beq
\label{DBImb}
\begin{array}{rcl}
{\cal L}_{DBI} &=& \Fint\bigint d^2\tilde\theta \Bigl[ ig{\cal Y}
\left( {\cal W} - {1\over2\kappa}\tilde\theta\tilde\theta \right)
-{1\over4} \xi_1\tilde\theta\tilde\theta\,{\cal W}
\crbig
&& \hspace{2.2cm}  + {1\over4}\Lambda U^2
-{1\over2} \Upsilon\left(U - {\cal W}+ {1\over2\kappa}\tilde\theta\tilde\theta\right) \Bigr] 
+ {\rm c.c.} + {\cal L}_{ST}.
\end{array}
\eeq
Both $U$ and $\Upsilon$ are unconstrained chiral $N=2$ superfields. The Lagrange multiplier
$\Upsilon$ imposes $U = {\cal W}-  {1\over2\kappa}\tilde\theta\tilde\theta$, which leads again to theory
(\ref{DBIa}). The first two terms, which have gauge and $N=2$ invariance properties related to 
the Maxwell character of ${\cal W}$ are left unchanged. The term quadratic in ${\cal W}$ has 
been turned into a linear one using the Lagrange multiplier. Hence,  the Maxwell superfield  
${\cal W}$, which contributes to Lagrangian (\ref{DBImb}) by
\beq
\label{DBImc}
\Fint\bigint d^2\tilde\theta\, {\cal W}\left( ig{\cal Y} + {1\over2}\Upsilon 
-{1\over4} \xi_1\,\tilde\theta\tilde\theta \right) + {\rm c.c.},
\eeq
can as well be eliminated: $\Upsilon$ should be such that this contribution is a derivative.
In terms of $N=1$ chiral superfields, ${\cal W}$ has components $X$ 
and $W_\alpha$ and since there exists two real superfields $V_1$ 
and $V_2$ such that $X={1\over2}\ov{DD}\, V_1$ and 
$W_\alpha = -{1\over4} \ov{DD}D_\alpha\, V_2$, we actually need to 
eliminate $V_1$ and $V_2$ with result
\beq
\label{DBImd}
\Upsilon = -i \widetilde{\cal W} - 2 ig{\cal Y} + {1\over2} (\xi_1 + i \zeta)\,
\tilde\theta\tilde\theta.
\eeq
In this expression, $\widetilde{\cal W}$ is a Maxwell $N=2$ superfield, the `magnetic dual' of
the eliminated ${\cal W}$.  There is a new 
arbitrary real deformation parameter $\zeta$, allowed by the field equation 
of $V_2$. Notice however that $\xi_1+i\zeta$ can be eliminated by a constant complex shift 
of $\Phi$. Invariance of $\Upsilon$ under the single-tensor gauge variation (\ref{chiral6})
implies that $\delta\widetilde{\cal W} = 2g \widehat{\cal W} =
-2g\delta{\cal Y}$ and
\beq
\label{DBIme}
{\cal Z } \equiv  \widetilde{\cal W} + 2g{\cal Y}
\eeq
is then a gauge-invariant chiral superfield. As already mentioned,
any unconstrained chiral $N=2$ superfield can be decomposed in this way and
our theory may as well be considered as a description of the chiral superfields ${\cal Z}$
and ${\cal Y}$ with Lagrangian
\beq
\label{DBImf}
{\cal L}_{DBI} =
\Fint\bigint d^2\tilde\theta \Bigl[ {1\over4}\Lambda U^2 
+ i U \Bigl({1\over2}{\cal Z} + {i\over4} (\xi_1+i\zeta)\tilde\theta\tilde\theta \Bigr)
+ {i\over4\kappa}\tilde\theta\tilde\theta ({\cal Z} - 2g {\cal Y}) \Bigr] + {\rm c.c.} + {\cal L}_{ST}.
\eeq
Invariance under the second supersymmetry implies that 
${\cal Z} + {i\over2}(\xi_1+i\zeta)\tilde\theta\tilde\theta$ transforms as a standard 
$N=2$ chiral superfield and then
\beq
\label{DBImg}
\delta^*_{deformed} \, {\cal Z} = i(\xi_1+i\zeta)\tilde\theta\eta 
+ i(\eta\tilde Q + \ov\eta\ov{\tilde Q}){\cal Z}.
\eeq
Eliminating $U$ leads finally to
\beq
\label{DBImh}
\widetilde{\cal L}_{DBI} =
\Fint\bigint d^2\tilde\theta \Bigl[ {1\over4\Lambda}
\Bigl({\cal Z} + {i\over2}(\xi_1+i\zeta)\tilde\theta\tilde\theta \Bigr)^2
+ {i\over4\kappa}\tilde\theta\tilde\theta ({\cal Z} - 2g {\cal Y}) \Bigr] + {\rm c.c.} + {\cal L}_{ST},
\eeq
which is the electric-magnetic dual of theory (\ref{DBIa}).\footnote{It reduces to Eq.~(\ref{DBIEM2})
if $g=0$.}
The Lagrange multiplier superfield $\Lambda^{-1}$ implies now the constraint
\beq
\label{DBImi}
0 = \left({\cal Z} + {i\over2}(\xi_1+i\zeta)\tilde\theta\tilde\theta \right)^2 
= {\cal Z}^2 + i(\xi_1+i\zeta)\tilde\theta\tilde\theta{\cal Z}.
\eeq
Using the expansion (\ref{chiral2b}),
$$
{\cal Z}(y,\theta,\tilde\theta) = Z(y,\theta) + \sqrt2\,\tilde\theta\omega(y,\theta)
-\tilde\theta\tilde\theta\left[ {i\over2}\Phi_{\cal Z}(y,\theta) + {1\over4} \ov{DD} \ov Z(y,\theta)\right],
$$
with $Z=\widetilde X + 2g Y$, $\omega_\alpha =  i\widetilde W_\alpha + 2g \chi_\alpha$
and $\Phi_{\cal Z}=2g\Phi$, this constraint corresponds to
$$
Z^2 = 0,
\qquad\qquad
Z\omega_\alpha =0,
\qquad\qquad
{1\over2}Z\ov{DD}\ov Z +\omega\omega =
-iZ[\Phi_{\cal Z} - (\xi_1+i\zeta)].
$$
In this case, and in contrast to the electric case, the constraint leading to the DBI theory
is due to the scale $\langle\Phi_{\cal Z}\rangle=2g\langle\Phi\rangle$: we will actually choose $\zeta=0$,
absorb $\xi_1$ into $\Phi_{\cal Z}$ and consider the constraint ${\cal Z}^2=0$ with a non-zero 
background value $\langle\Phi_{\cal Z}\rangle$ breaking the second supersymmetry.
Our magnetic theory is then
\beq
\label{DBImk}
\widetilde{\cal L}_{DBI} =
\Fint\bigint d^2\tilde\theta \Bigl[ {1\over4\Lambda}
{\cal Z}^2
+ {i\over4\kappa}\tilde\theta\tilde\theta ({\cal Z} - 2g {\cal Y}) \Bigr] + {\rm c.c.} + {\cal L}_{ST},
\eeq
with constraints
\beq
\label{DBImj}
Z^2 = 0,
\qquad\qquad
Z\omega_\alpha =0,
\qquad\qquad
{1\over2}Z\ov{DD}\ov Z +\omega\omega =
-iZ\Phi_{\cal Z},
\eeq
the DBI scale arising from $\Phi_{\cal Z} = \phi_{\cal Z} + \langle\Phi_{\cal Z}\rangle$.
As in the Maxwell case, the third equation, which also reads
\beq
\label{DBImk2}
Z = {i\omega\omega \over \Phi_{\cal Z}  - {i\over2}\ov{DD}\ov Z},
\eeq
implies $Z\omega_\alpha = Z^2=0$ and allows to express $Z$ as a function of $\omega\omega$
and $\Phi$, $Z=Z(\omega\omega,\Phi)$, using $\Phi_{\cal Z} = 2g\Phi - \xi_1$. 
The magnetic theory (\ref{DBImk}) is then simply
\beq
\label{DBIml}
\widetilde{\cal L}_{DBI} = -{1\over2\kappa} \Im
\Fint \Bigl[Z(\omega\omega,\Phi) - 2g Y \Bigr] + {\cal L}_{ST}.
\eeq
It is the electric-magnetic dual of expression (\ref{DBIb}).
At this point, it is important to recall that $\omega$ and $\Phi$ are actually $N=1$ superfields
components of ${\cal Z} = \widetilde{\cal W} + 2g{\cal Y}$, {\it i.e.}
\beq
\omega_\alpha = i\widetilde W_\alpha + 2g\chi_\alpha.
\eeq
The kinetic terms for the single-tensor multiplet $(L,\Phi)$, $L=D\chi-\ov D\ov\chi$, are included in 
${\cal L}_{ST}$ while $Z(\omega\omega,\Phi)$ includes the DBI kinetic terms for the
Maxwell $N=1$ superfield $\widetilde W_\alpha$. As in the electric case, the magnetic theory has a 
contribution proportional to the four-form field included in $Y$. 

The third constraint (\ref{DBImj}) is certainly invariant under the variations (\ref{chiral2c}), using 
$Z\omega_\alpha=0$. But with a non-zero background value  $\Phi = \phi + \langle\Phi\rangle$,
the spinor $\omega_\alpha$ transforms nonlinearly, like a goldstino:\footnote{See Eq.~(\ref{DBImg}).}
\beq
\label{DBImm}
\delta^*\omega_\alpha = -{i\over\sqrt2} \langle\Phi\rangle\,\eta_\alpha
-{i\over\sqrt2} \phi\,\eta_\alpha - {\sqrt2\over4} \eta_\alpha \, \ov{DD}\, \ov Z
-\sqrt2 i (\sigma^\mu\ov\eta)_\alpha \partial_\mu Z.
\eeq
The solution of the constraint (\ref{DBImk2}) is given in Appendix B. The bosonic Lagrangian included 
in the magnetic theory (\ref{DBIml}) is 
\beq
\label{DBIbos1}
\begin{array}{rcl}
\widetilde{\cal L}_{DBI,bos.} &=& {\Re\Phi_{\cal Z}\over8\kappa} 
- {\Re\Phi_{\cal Z}\over8\kappa|\Phi_{\cal Z}|^2}
\Biggl\{ - |\Phi_{\cal Z}|^4 \,{\rm det} \left[\eta_{\mu\nu} - 2\sqrt2\,|\Phi_{\cal Z}|^{-1} 
(\widetilde F_{\mu\nu} - gb_{\mu\nu}) \right]
\crbig
&&
- 8 \tilde d{_2}^2 \, (|\Phi_{\cal Z}|^2 + 2g^2C^2) + 2 g^2 C^2 |\Phi_{\cal Z}|^2
\crbig
&& + 8gC\tilde d_2 \, \epsilon^{\mu\nu\rho\sigma}
(\widetilde F_{\mu\nu} - g \, b_{\mu\nu})(\widetilde F_{\rho\sigma} - g \, b_{\rho\sigma})
\Biggr\}^{1/2}
\crbig
&& - {\Im\Phi_{\cal Z}\over8\kappa|\Phi_{\cal Z}|^2} \left[ \epsilon^{\mu\nu\rho\sigma}
(\widetilde F_{\mu\nu} - g \, b_{\mu\nu})(\widetilde F_{\rho\sigma} - g \, b_{\rho\sigma})
- 4gC \widetilde d_2 \right]
\crbig
&&
+ {g\over24\kappa}\epsilon^{\mu\nu\rho\sigma} C_{\mu\nu\rho\sigma} + {\cal L}_{ST,bos.}.
\end{array}
\eeq
It depends on a single auxiliary field, the Maxwell real scalar $\widetilde d_2$, with field equation
\beq
\begin{array}{rcl}
\widetilde d_{2, \, bos.} &=& \displaystyle-\frac{g\,C}{2(|\Phi_{\cal Z}|^2+2g^2C^2)} \,
\epsilon^{\mu\nu\rho\sigma} 
(\widetilde F_{\mu\nu} - g \, b_{\mu\nu})(\widetilde F_{\rho\sigma} - g \, b_{\rho\sigma})
\crbig
&& \displaystyle
- \frac{g\,C\Im\Phi_{\cal Z}}{2|\Phi_{\cal Z}|^2}
\frac{\sqrt{-\det\Bigl(\eta_{\mu\nu}+\frac{2\sqrt2}{\sqrt{2g^2C^2+|\Phi_{\cal Z}|^2}}
(\widetilde F_{\mu\nu} - g \, b_{\mu\nu})\Bigr)}}
{\sqrt{(\Re\Phi_{\cal Z})^2+2g^2C^2}} \, .
\end{array}
\eeq
Eliminating $\tilde d_2$ and using $\Phi_{\cal Z}=2g\Phi - \xi_1$ to reintroduce the superfield 
$\Phi$ of the single-tensor multiplet and the `original' Fayet-Iliopoulos term $\xi_1$, we 
finally obtain the magnetic, bosonic Lagrangian
\beq
\label{DBIbos2}
\begin{array}{rcl}
\widetilde{\cal L}_{DBI,bos.} &=&  \displaystyle{2g\Re\Phi - \xi_1 \over 8\kappa} 
-{1\over8\kappa}\sqrt{(2g\Re\Phi -\xi_1)^2+ 2 g^2 C^2 }
\crbig
&& \hspace{2.3cm} \times \sqrt{-\det\Big(\eta_{\mu\nu}
-\frac{2\sqrt2}{\sqrt{2g^2 C^2 + |2g\Phi - \xi_1|^2}}(\widetilde F_{\mu\nu} - g b_{\mu\nu})\Big)}\Biggr)
\crbig
&& \displaystyle -{g\Im\Phi \over 4\kappa(2g^2C^2 
+ |2g\Phi - \xi_1|^2)}\epsilon^{\mu\nu\rho\sigma}(\widetilde F_{\mu\nu} - g b_{\mu\nu}) 
(\widetilde F_{\rho\sigma} - g b_{\rho\sigma})
\crbig
&& \displaystyle + {g\over24\kappa}\epsilon^{\mu\nu\rho\sigma} C_{\mu\nu\rho\sigma}
+ {\cal L}_{ST,bos.} \,.
\end{array}
\eeq
As in the electric case, the DBI term has a field-dependent coefficient,
\beq
- {1\over8\kappa} \sqrt{(2g\Re\Phi - \xi_1)^2 + 2g^2 C^2 } \,
\sqrt{-\det \Big(\eta_{\mu\nu}-\frac{1}{\sqrt{ 2g^2C^2 +|2g\Phi - \xi_1|^2}}
(\widetilde F_{\mu\nu}-gb_{\mu\nu})\Big)},
\eeq
and, as expected, the scalar potentials of the magnetic and electric [Eq.~(\ref{DBIpot})] theories
are identical.

Define the complex dimensionless field
\beq
\label{DBIbos3}
S =  \kappa\sqrt{(2g\Re\Phi - \xi_1)^2 + 2g^2 C^2 } +2i\kappa g\Im\Phi,
\eeq
for which $\kappa^{-2} |S|^2 = |2g\Phi-\xi_1|^2 + 2g^2C^2$. In terms of $S$, the magnetic theory
(\ref{DBIbos2}) rewrites as
\beq
\label{S1}
\begin{array}{rcl}
\widetilde{\cal L}_{DBI,bos.} &=& \displaystyle{2g\Re\Phi - \xi_1 \over 8\kappa} 
-{1\over8\kappa^2} \Re{1\over S} \sqrt{-\det\Bigl( |S|\eta_{\mu\nu} 
- 2\sqrt2\kappa (\widetilde F_{\mu\nu} - g b_{\mu\nu})\Bigr)}
\crbig
&& \displaystyle + {1\over8} \Im {1\over S}\,
\epsilon^{\mu\nu\rho\sigma}(\widetilde F_{\mu\nu} - g b_{\mu\nu}) 
(\widetilde F_{\rho\sigma} - g b_{\rho\sigma})
+ {g\over24\kappa}\epsilon^{\mu\nu\rho\sigma} C_{\mu\nu\rho\sigma}
+ {\cal L}_{ST,bos.}
\crbig
&=& \displaystyle{2g\Re\Phi - \xi_1 \over 8\kappa} 
-{1\over8\kappa^2} \Re S \sqrt{-\det\Bigl( \eta_{\mu\nu} 
- 2\sqrt2\kappa |S|^{-1}(\widetilde F_{\mu\nu} - g b_{\mu\nu})\Bigr)}
\crbig
&& \displaystyle + {1\over8} \Im {1\over S}\,
\epsilon^{\mu\nu\rho\sigma}(\widetilde F_{\mu\nu} - g b_{\mu\nu}) 
(\widetilde F_{\rho\sigma} - g b_{\rho\sigma})
+ {g\over24\kappa}\epsilon^{\mu\nu\rho\sigma} C_{\mu\nu\rho\sigma}
+ {\cal L}_{ST,bos.} .
\end{array}
\eeq
This is to be compared with the electric theory (\ref{DBId}):
\beq
\label{S2}
\begin{array}{rcl}
{\cal L}_{DBI,\, bos.} &=&  \displaystyle {2g\Re\Phi - \xi_1 \over 8\kappa} - {1\over8\kappa^2} \Re S
\sqrt{-\det (\eta_{\mu\nu}-2\sqrt 2\kappa\,F_{\mu\nu})}
\crbig
&& \displaystyle + {1\over8}\Im S \, \epsilon^{\mu\nu\rho\sigma}F_{\mu\nu}F_{\rho\sigma}
- {g\over4}\epsilon^{\mu\nu\rho\sigma}b_{\mu\nu}F_{\rho\sigma} 
+ {g\over24\kappa} \epsilon^{\mu\nu\rho\sigma}C_{\mu\nu\rho\sigma} 
+ {\cal L}_{ST, \, bos.}.
\end{array}
\eeq
Hence, the duality from the electric to the magnetic theory corresponds to the transformations
\beq
\label{S3}
b_{\mu\nu} \,\rightarrow\, 0, \qquad
F_{\mu\nu} \,\rightarrow\, \widetilde F_{\mu\nu} - g b_{\mu\nu}, \qquad
S\,\rightarrow\, S^{-1}, \qquad
\eta_{\mu\nu}\,\rightarrow\, |S|\eta_{\mu\nu},
\eeq
which can be also derived from electric-magnetic duality applied on the bosonic DBI theory
only.

%*******************************************
\subsection{Double-tensor formulation and connection with the string fields}
%*******************************************

As mentioned in the introduction, in IIB superstrings compactified to four dimensions with eight 
residual supercharges, the dilaton belongs to a double-tensor supermultiplet. 
This representation of $N=2$ supersymmetry includes 
two Majorana spinors, two antisymmetric tensors $B_{\mu\nu}$ (NS--NS) and $C_{\mu\nu}$ (R--R)
with gauge symmetries
\beq
\label{antisymsym}
\delta_{gauge}\,B_{\mu\nu} = 2\,\partial_{[\mu} \Lambda_{\nu]}, \qquad\qquad
\delta_{gauge}^{\,\prime} \,C_{\mu\nu} = 2\,\partial_{[\mu}\Lambda^\prime_{\nu]}
\eeq
and two (real) scalar fields, the NS--NS dilaton and the R--R scalar, for a total of $4_B+4_F$ physical
states. In principle, both antisymmetric tensors can be dualized to pseudoscalar fields with 
axionic shift symmetry, in a version of the effective field theory where the dilaton belongs to a 
hypermultiplet with four scalars in a quaternion-K\"ahler\footnote{For supergravity. The limit of
global supersymmetry is a hyperk\"ahler manifold, which is Ricci-flat.}
manifold possessing three perturbative shift isometries, since the R--R scalar 
has its own shift symmetry. It is easy to see that only two shift isometries, related to the two 
antisymmetric tensors, commute, 
while all three together form the Heisenberg algebra. Indeed, in the double-tensor basis, the R--R field 
strength is modified~\cite{IIBsugra} due to its anomalous Bianchi identity to 
$3\,\partial_{[\lambda}C_{\mu\nu]}-3\,C^{(0)}\partial_{[\lambda}B_{\mu\nu]}$. Thus, a shift of 
the R--R scalar $C^{(0)}$ by a constant $\lambda$ is accompanied by an appropriate 
transformation of $C_{\mu\nu}$ to leave its modified field-strength invariant:
\beq
\label{Heisenberg}
\delta_{H}C^{(0)}=\lambda, \qquad \delta_H C_{\mu\nu}=\lambda B_{\mu\nu} .
\eeq
It follows that $\delta_{gauge}$, $\delta_{gauge}'$ and $\delta_H$ verify the Heisenberg algebra, 
with a single non-vanishing commutator 
\beq
\label{commutator}
\left[\delta_{gauge},\delta_H\right]=\delta_{gauge}^{\,\prime}\, .
\eeq

To establish the connection of the general formalism described in the previous subsections with 
string theory, we would like to identify the double-tensor multiplet with the universal dilaton 
hypermultiplet and study its coupling to the Maxwell goldstino multiplet of a single D-brane, in the 
rigid (globally-supersymmetric) limit. To this end, we transform the $N=2$ double-tensor into a 
single-tensor representation by dualizing one of its two $N=1$ linear multiplet components $L'$, 
containing the R--R fields $C_{\mu\nu}$ and $C^{(0)}$, into a chiral basis 
$\Phi+\ov{\Phi}$. In this basis, the two R--R isometries correspond to constant complex shifts of the 
$N=1$ superfield $\Phi$. Imposing this symmetry to the kinetic function of 
Eqs.~(\ref{ST7})--(\ref{ST8}), one obtains (up to total derivatives, after superspace integration):
\beq
\label{kinetic}
{\cal H}(L,\Phi,\ov{\Phi})=\alpha\Bigl(-{1\over 3}L^3 + {1\over 2}L (\Phi+\ov{\Phi})^2\Bigr)
+\beta\Bigl(-L^2+{1\over2}(\Phi+\ov{\Phi})^2\Bigr)\, ,
\eeq
where $\alpha$ and $\beta$ are constants. Note that the second term proportional to $\beta$ can be obtained from the first by shifting $L+\beta/\alpha$. For $\alpha=0$ however, it corresponds to the free case of quadratic kinetic terms for all fields of the single-tensor multiplet.
The coupling to the Maxwell goldstino multiplet is easily obtained 
using Eqs.~(\ref{DBIc}), (\ref{V1WWis}) and (\ref{CS1}). Up to total derivatives, 
the action is:
\beq
\label{DT1}
\begin{array}{rcl}
{\cal L} &= &\Dint\Big[\alpha\Big(-{1\over 3}L^3 + {1\over 2}L (\Phi+\ov{\Phi})^2\Big)
+\beta\Big(-L^2+{1\over2}(\Phi+\ov{\Phi})^2\Big)
\crbig
&&\hspace{.4cm}
-g(\Phi+\ov{\Phi})V_1(WW)\Big]
+g\Fint \Bigl[\chi^\alpha W_\alpha - {i\over 2 \kappa} Y -\frac{\xi_1}{4g}X(WW) \Bigr] + {\rm c.c.}
\end{array}
\eeq
In general, the four-form field is not inert under the variation $\delta_H$ of Eq.~(\ref{Heisenberg}) \cite{BT}.
In our single-tensor formalism, $\delta_H L=0$ and $\delta_H\Phi = c$ where $c$ is complex when combined with the axionic shift  $\delta'_{gauge}$ of ${\rm Im}\Phi$ dual to $C_{\mu\nu}$ of Eq.~(\ref{antisymsym}); in addition
\beq
\label{Yvar}
\delta_H Y = - ic\kappa X(WW).
\eeq
With this variation, the Lagrangian, including the Chern-Simons interaction, is invariant under 
the Heisenberg symmetry.

We can now dualize back $\Phi+\ov{\Phi}$ to a second linear multiplet $L'$ by first
replacing it with a real superfield $U$:
\beq
\label{DT2}
\begin{array}{rl}
{\cal L} = &\Dint\Big[\alpha\left(-{1\over 3}L^3 + {1\over 2}L U^2\right)
+\beta\left(-L^2+{1\over2}U^2\right)-U (mL'+ gV_1)\Big]
\crbig
&+g\Fint \Bigl[ \chi^\alpha W_\alpha - {i\over 2 \kappa} Y -\frac{\xi_1}{4g}X \Bigr] + {\rm c.c.},
\end{array}
\eeq
where the constant $m$ corresponds to a rescaling of $L'$. Solving for $U$,
\beq
\label{DT3}
U={mL'+gV_1\over \alpha L+\beta}\,,
\eeq
delivers the double-tensor Lagrangian
\beq
\label{DT4}
{\cal \widetilde L} =
\Dint\Big[ -{\alpha\over 3}L^3 -\beta L^2- {1\over2}{(mL'+gV_1)^2\over \alpha L
+\beta}\Big] 
+g\Fint \Bigl[ \chi^\alpha W_\alpha - {i\over 2 \kappa} Y - {\xi_1\over 4g}X \Bigr] + {\rm c.c.},
\eeq
where as before $V_1= V_1(WW)$ and $X=X(WW)= {1\over2}\ov{DD}\,V_1(WW)$. 
It is invariant under variation (\ref{Yvar}) of the four-form superfield combined with
$\delta_H L^\prime = 2c(\alpha L+\beta)/m$.

After elimination of the Maxwell auxiliary field (choosing $m=\sqrt2$)
\beq
d_{2, \,bos.} = {gC\over2\kappa}  \sqrt{-\det (\eta_{\mu\nu}+2\sqrt 2\kappa\,F_{\mu\nu}) \over 
\left({\sqrt2 g\,C^\prime\over \alpha C+\beta} - \xi_1\right)^2 + 2g^2 C^2} \, ,
\eeq
the component expansion of the bosonic Lagrangian is 
\beq
\label{DT8}
\begin{array}{rcl}
{\cal \widetilde L}_{bos.} &=& (\alpha C+\beta)\left[{1\over 2} (\partial_\mu C)^2+{1\over 2}
\partial_\mu\Big({C'\over \alpha C+\beta}\Big)^2 +{1\over 12} (3 \,\partial_{[\mu} b_{\nu\rho]})^2\right]  
\crbig
&&  +{1\over 12(\alpha C+\beta)}\left(3\,\partial_{[\mu} b'_{\nu\rho]}
+{g\kappa\over \sqrt{2}}\omega_{\mu\nu\rho}
-{C'\over \alpha C + \beta} 3\, \partial_{[\mu} b_{\nu\rho]}\right)^2 
\crbig
&& -{g\over 4 \kappa\sqrt2}({C'\over \alpha C+\beta}+{\xi_1\over\sqrt2 g})+ {g\over 4\kappa\sqrt{2}}
\sqrt{({C'\over \alpha C+\beta}+{\xi_1\over \sqrt2g})^2+C^2}
\sqrt{-\det(\eta_{\mu\nu}+2\sqrt{2}\kappa F_{\mu\nu})}
\crbig
&&- \frac{g}{4} \epsilon^{\mu\nu\rho\sigma} b_{\mu\nu} F_{\rho\sigma}
+{g\over 24\kappa}\epsilon^{\mu\nu\rho\sigma}C_{\mu\nu\rho\sigma} \,.
\end{array}
\eeq
in terms of the Maxwell Chern-Simons form $\omega_{\nu\rho\sigma} = 3\, A_{[\nu} F_{\rho\sigma]}$.

We expect that this action describes the globally-supersymmetric limit of the effective four-dimensional 
action of a D-brane coupled to the universal dilaton hypermultiplet of the perturbative type II string. 
As mentioned previously, its general form in the local case depends also on two constant parameters, 
upon imposing the perturbative Heisenberg isometries, that correspond to the tree and one-loop 
contributions~\cite{oneloop}. It is tempting to identify these two parameters with $\alpha$ and 
$\beta$ of our action. Moreover, by identifying the two antisymmetric tensors 
$b_{\mu\nu}$ and $b^\prime_{\mu\nu}$ with the respective NS--NS $B_{\mu\nu}$ and R--R 
$C_{\mu\nu}$ and the combination $C'/(\alpha C+\beta)$ with the R--R 
scalar $C^{(0)}$, as the Heisenberg transformations indicate, one finds that the two actions match 
up to normalization factors depending on the NS--NS dilaton that should correspond to the scalar 
$C$. Finding the precise identifications, which certainly depend on the way one 
should take the rigid limit that decouples gravity, is an interesting question beyond our 
present analysis restricted to global supersymmetry.

%%%%%%%%%%%%
\section{Nonlinear \boldmath{$N=2$} QED}  \label{secQED}
%%%%%%%%%%%%

We will now show that the effective theory presented above describing a super-Higgs 
phenomenon of partial (global) supersymmetry breaking can be identified with the Higgs 
phase of nonlinear $N=2$ QED, up to an appropriate choice of the single-tensor multiplet 
kinetic terms. We will then analyze its vacuum structure in the generally allowed parameter 
space.

In linear $N=2$ quantum electrodynamics (QED), the Lagrangian couples a hypermultiplet with 
two chiral superfields $(Q_1,Q_2)$ to the vector multiplet $(V_1,V_2)$ or $(X,W_\alpha)$. 
The $U(1)$ gauge transformations 
of the hypermultiplet are linear, and $Q_1$ and $Q_2$ have opposite $U(1)$ charges:
\beq
\label{QED1}
{\cal L}_{QED} = \Dint \left[ \ov Q_1Q_1 e^{V_2} + \ov Q_2Q_2 e^{-V_2} \right]
+ \Fint {i\over\sqrt2} XQ_1Q_2 + {\rm c.c.} + {\cal L}_{Max.} + \Delta{\cal L},
\eeq
where ${\cal L}_{Max.}$ includes (canonical) gauge kinetic terms and $\Delta{\cal L}$ 
contains three parameters:
\beq
\label{QED1b}
\Delta{\cal L} = m\Fint Q_1Q_2 + {\rm c.c.} + \Dint [\xi_1V_1 + \xi_2V_2].
\eeq
The hypermultiplet mass term with coefficient $m$ can be eliminated by a shift of $X$ and 
$\xi_{1,2}$ are the two Fayet-Iliopoulos coefficients. 
Since $\xi_1\int d^2\theta d^2\ov\theta\, V_1 = -{1\over4}\int d^2\theta\, \xi_1X + {\rm c.c.}$, the 
complete superpotential $w$ is
$$
w = \left({i\over\sqrt2} X + m\right) Q_1Q_2 - {1\over4}\xi_1X.
$$
There are six real auxiliary fields, $f_{Q_1}$, $f_{Q_2}$, $d_1$ and $d_2$ but only four are actually independent:\footnote{
We use the same notation for a chiral superfield $\Phi$, $Q_1$, $Q_2$, \dots and for its
lowest complex scalar component field.} $Q_1 \ov f_{Q_1} = Q_2 \ov f_{Q_2}$.
Since the metric is canonical, $\det K_{i\ov j} = 1$ and trivially hyperk\"ahler.
If $\xi_1=\xi_2=0$, the gauge symmetry is not broken and the hypermultiplet mass 
$m + i \langle X\rangle / \sqrt2 $ is arbitrary.
Any nonzero $\xi_1$ or $\xi_2$ induces $U(1)$ symmetry breaking with all fields having the same mass. In any case, $N=2$ supersymmetry remains unbroken at the global minimum.

In order to first bring the theory to a form allowing dualization to our single-tensor formulation,
we use the holomorphic field redefinition\footnote{This field redefinition  
has constant Jacobian.}
\beq
\label{QED2}
\begin{array}{c}
Q_1 = a\, \sqrt\Phi \, e^{\Phi^\prime}, \qquad  \qquad
Q_2 = ia\, \sqrt\Phi \, e^{-\Phi^\prime} ,
\crbig
Q_1Q_2 = ia^2\Phi , \qquad  \qquad Q_1/Q_2 = -ie^{2\Phi^\prime} ,
\end{array}
\eeq
with $a^2 = 1/\sqrt2$. The QED Lagrangian becomes
\beq
\label{QED3}
\begin{array}{rcl}
{\cal L}_{QED} &=& {1\over\sqrt2} \Dint 
\sqrt{\Phi\ov\Phi}  \left[ e^{\Phi^\prime+\ov\Phi^\prime + V_2} 
+ e^{-\Phi^\prime-\ov\Phi^\prime - V_2}\right]
+ {\cal L}_{Max.} 
\crbig
&& 
+ \Fint \left[ -{1\over2} \Phi (X-\sqrt2im) - {1\over4}\xi_1 X \right] + {\rm c.c.} 
+ \xi_2\Dint V_2.
\end{array}
\eeq
While the gauge transformation of $\Phi^\prime$ is $\delta_{U(1)}\Phi^\prime = \Lambda_c$, 
$\Phi$ is gauge invariant. Since the K\"ahler potential is now a function of 
$\Phi^\prime+\ov\Phi^\prime$, with a St\"uckelberg gauging of the axionic shift
of $\Phi^\prime$, the chiral $\Phi^\prime$ can be dualized to a linear $L$ using a $ N=1$ Legendre
transformation. The result is
\beq
\label{QED5}
\begin{array}{rcl}
{\cal L}_{QED} &=& \Dint 
\left[ \sqrt{2\Phi\ov\Phi + L^2} - L \ln\left( \sqrt{2\Phi\ov\Phi + L^2} + L\right) 
\right] + {\cal L}_{Max.}
\crbig
&& -  \Fint \left[{1\over2}X\Phi + \chi^\alpha W_\alpha 
- {i\over\sqrt2}m\Phi + {1\over4}\xi_1X \right] + {\rm c.c.} + \xi_2\Dint V_2.
\end{array}
\eeq
The dual single-tensor QED theory has off-shell $N=2$ invariance (the Laplace equation (\ref{ST7}) 
is verified) and the two multiplets are now coupled by a $N=2$ Chern-Simons interaction 
(\ref{CS3}). Notice that the free quadratic kinetic terms of the charged hypermultiplet lead to a 
highly non-trivial kinetic function in the single-tensor representation. Moreover, 
there are only four auxiliary fields, $f_\Phi$, $d_1$ and $d_2$.
The Legendre transformation defines the scalar field $C$ in $L$ as
\beq
e^{2\Re\Phi^\prime} = {1\over\sqrt{2\Phi\ov\Phi}}\left( \sqrt{2\Phi\ov\Phi+C^2} + C \right),
\qquad
e^{-2\Re\Phi^\prime} = {1\over\sqrt{2\Phi\ov\Phi}}\left( \sqrt{2\Phi\ov\Phi+C^2} - C \right)
\eeq
and Eqs.~(\ref{QED2}) relate then $C$ and $\Phi$ with $Q_1$ and $Q_2$:
\beq
\label{QED8}
C = |Q_1|^2 - |Q_2|^2, \qquad\qquad \Phi = -\sqrt2i\,Q_1Q_2.
\eeq

According to Eq.~(\ref{DBIb}), the nonlinear DBI version of $N=2$ QED is obtained by replacing
in Lagrangian (\ref{QED5}) $X$ by $X(WW)$, which includes DBI gauge kinetic terms, by
omitting ${\cal L}_{Max.}$ which is removed by the third constraint (\ref{DBI6}) and by adding the 
four-form term ${i\over2\kappa}\int d^2\theta\, Y + {\rm c.c.}$:
\beq
\label{QED6}
\begin{array}{rcl}
{\cal L}_{QED, DBI} &=& \Dint 
\left[ \sqrt{2\Phi\ov\Phi + L^2} - L \ln\left( \sqrt{2\Phi\ov\Phi + L^2} + L\right)
+ \xi_2\, V_2 \right] 
\crbig
&& -\Fint\left[ \left({1\over2}\Phi + {1\over4}\xi_1 \right)  X(WW) - {i\over\sqrt2}m\Phi 
+ \chi^\alpha W_\alpha - {i\over2\kappa} Y \right]
+ {\rm c.c.} 
\end{array}
\eeq
Notice that two additional terms appear compared to the action studied in Section~\ref{secDBI}: an 
Fayet-Iliopoulos term proportional to $\xi_2$ and a term linear in $\Phi$ which is also invariant 
under the second (nonlinear) supersymmetry (\ref{ST4}); they generate, together with $\xi_1$ 
the general parameter space of nonlinear QED coupled to a charged hypermultiplet.
Without loss of generality, we choose $m$ to be real, while
the choice $\xi_1 = -1/\kappa$ would canonically normalize gauge kinetic terms for a background 
where $\Phi$ vanishes.
We may return to chiral superfields $(\Phi,\Phi^\prime)$ or $(Q_1,Q_2)$ to write the DBI theory
as\footnote{See Eq.~(\ref{hyp1}).}
\beq
\label{QED4}
\begin{array}{rcl}
{\cal L}_{QED} &=& \Dint \left[ \ov Q_1Q_1 e^{V_2} + \ov Q_2Q_2 e^{-V_2} + \xi_2 V_2\right]
\crbig
&&
+\Fint\left[ \left( {i\over\sqrt2}Q_1Q_2 - {1\over4}\xi_1 \right)  X(WW) + mQ_1Q_2 
+ {i\over2\kappa} Y \right]
+ {\rm c.c.} 
\end{array}
\eeq
Since $X(WW)|_{\theta=0}$ only depends on fermion fields, 
the auxiliary fields $f_1$ and $f_2$ only contribute to
the bosonic Lagrangian by a hypermultiplet mass term
$$
\Bigl( |f_1|^2 + |f_2|^2 \Bigr)_{bos.}= m^2 \left( |Q_1|^2 + |Q_2|^2 \right) 
$$
to be added to the scalar potential obtained from Eq.~(\ref{DBIpot}) with the substitutions
$$
2 g\Re\Phi-\xi_1\,\longrightarrow\, 2\sqrt2\Im (Q_1Q_2)-\xi_1, \qquad\qquad
gC \,\longrightarrow\, C+\xi_2 = \xi_2 + |Q_1|^2 - |Q_2|^2
$$
(since we have chosen $g=1$). The complete potential is then\footnote{The auxiliary $d_2$ is given 
in Eq.~(\ref{d2elec}).}
\beq
\begin{array}{rcl}
V_{QED,DBI} &=& \displaystyle{1\over8\kappa}\left(2\sqrt2\Im (Q_1Q_2)-\xi_1 \right) \left[ 
\sqrt{1 + { 2[\xi_2 + |Q_1|^2 - |Q_2|^2]^2
\over [2\sqrt2\Im (Q_1Q_2)-\xi_1]^2}} - 1 \right]
\crbig
&& + m^2 \left( |Q_1|^2 + |Q_2|^2 \right). 
\end{array}
\eeq
The analysis is then very simple. The first line vanishes only for 
\beq
\label{vac1}
\langle \xi_2 + |Q_1|^2 - |Q_2|^2 \rangle = 0, \qquad\qquad
\langle 2\sqrt2\Im (Q_1Q_2)-\xi_1\rangle > 0.
\eeq
The first condition is the usual $D$--term equation $\langle d_2\rangle =0$ for the Maxwell superfield.  
The second condition is necessary to have a well-defined DBI gauge kinetic term at the minimum. 
Hence, if $m=0$, conditions (\ref{vac1}), which can always be solved, define the vacuum of the theory. 
Choosing $\langle Q_1 \rangle = v$ and $\langle Q_2 \rangle=\sqrt{v^2+\xi_2}$, with $v$ real
(and arbitrary), we find 
a massive vector boson which, along with a real scalar and the two Majorana fermions
$$
{1\over \sqrt{2 v^2+\xi_2}} \, \left[ v \psi_{Q_1}- \sqrt{v^2+\xi_2}\, \psi_{Q_2} 
\right] \pm i\lambda ,
$$ 
makes a massive $N=1$ vector multiplet of mass $\sqrt{v^2+\xi_2/2}$. Hence the potentially 
massless gaugino $\lambda$, with its goldstino-like second supersymmetry variation 
$\delta^*\lambda_\alpha = - {1\over\sqrt2\kappa}\eta_\alpha + \ldots$, has been absorbed 
in the massive 
$U(1)$ gauge boson multiplet. This is possible only because the second supersymmetry 
transformation  of the four-form field compensates the gaugino nonlinear variation.
The fermion
$$
\sqrt{v^2 + \xi_2} \, \psi_{Q_1} + v \, \psi_{Q_2}
$$
is massless and corresponds to the fermion of the chiral superfield $\Phi$ in the single-tensor 
formalism, in agreement with our analysis in Section~\ref{couplingDBIst} [see below 
Eq.~(\ref{DBIpot})]. With two real scalars, it belongs to a massless $N=1$ chiral multiplet.

If $m\ne0$, a supersymmetric vacuum has $\langle Q_1 \rangle = \langle Q_2 \rangle
=0$. It only exists if $\xi_2=0$ and $\xi_1\ne0$. The second condition is again to have DBI gauge 
kinetic terms on this vacuum. In this case, the $U(1)$ gauge symmetry is not broken, the goldstino 
vector multiplet remains massless and the hypermultiplet has mass $m$. If $m\ne0$, a nonzero 
Fayet-Iliopoulos coefficient $\xi_2$ breaks then $N=1$ linear supersymmetry. Note that the 
single-tensor formalism is appropriate for the description of the Higgs phase of nonlinear QED in a 
manifest $N=1$ superfield basis (with respect to the linear supersymmetry), while the charged 
hypermultiplet representation is obviously convenient for describing the Coulomb phase. 

One can finally expand the action (\ref{QED4}) in powers of $\kappa$ in order to find the lowest 
dimensional operators that couple the goldstino multiplet of partial supersymmetry breaking to the 
$N=2$ hypermultiplet. Besides the dimension-four operators corresponding to the gauge factors 
$e^{\pm V_2}$, one obtains a dimension-six superpotential interaction $\sim\kappa Q_1Q_2W^2$ 
coming from the solution of the nonlinear constraint $X=\kappa W^2+{\cal O}(\kappa^3)$; it amounts 
to a field-dependent correction to the $U(1)$ gauge coupling.

%**************************************************
\section{Conclusions} \label{secfinal}
\setcounter{equation}{0}
%**************************************************

In this work we have studied the interaction of the Maxwell goldstino multiplet of $N=2$ nonlinear 
supersymmetry to a hypermultiplet with at least one isometry. The starting point was to describe the 
hypermultiplet in terms of a single-tensor multiplet, which admits an off-shell $N=2$ formulation, and 
introduce a coupling using a Chern-Simons interaction. This system describes the coupling of a 
D-brane to bulk fields of $N=2$ compactifications of type II strings, in the rigid limit of decoupled 
gravity. Using $N=1$ and $N=2$ dualities, we have also obtained equivalent 
formulations of the nonlinear Maxwell theory coupled to a matter $N=2$ supermultiplet. This web
of theories is summarized in the Figure. 

\begin{figure}[htbp]
\label{figure}
\begin{center}
\setlength{\unitlength}{1cm}%
\begin{picture}(15,7.5)(-0.5,0.5)

\thicklines

\put(0.1,4.1){Single-tensor}
\put(0.1,3.6){St\"uckelberg}
\put(0.1,3.1){gauging}
\put(0.1,2.5){$(L',\Phi')$ (\ref{STprime})}

\put(3.0,3.4){\vector(1,0){2.5}}
\put(5.5,3.6){\vector(-1,0){2.5}}
\put(3.0,4){ST-ST duality}

\put(6,4.1){Single-tensor}
\put(6,3.6){Chern-Simons}
\put(6,3){$(L,\Phi)$ \,\, (\ref{DBIb})}

\put(9.0,3.6){\vector(1,0){2.5}}
\put(11.5,3.4){\vector(-1,0){2.5}}
\put(9.2,4){E-M duality}

\put(12,4.1){Magnetic dual}
\put(12,3.6){Single-tensor}
\put(12,3){$(L,\Phi)$ \quad (\ref{DBIml})}

\put(7.2,5.2){\vector(0,1){0.8}}
\put(7.2,5.8){\vector(0,-1){0.8}}

\put(6,6.8){Double-tensor}
\put(6,6.2){$(L,L')$ \,\, (\ref{DT})}

\put(7.2,1.5){\vector(0,1){1}}
\put(7.2,2.5){\vector(0,-1){1}}

\put(6,1.1){Hypermultiplet}
\put(6,0.5){$(\Phi,\Phi')$ \quad (\ref{hyp1})}

\end{picture}%
\end{center}
\begin{quote} 
\caption{Web of dualities: double arrows indicate duality transformations preserving off-shell 
$N=2$  supersymmetry, simple arrows are $N=1$ off-shell dualities only, leading to theories 
with on-shell $N=2$ supersymmetry. The $N=1$ superfields and the related equations
are indicated. }
\end{quote}
\end{figure}
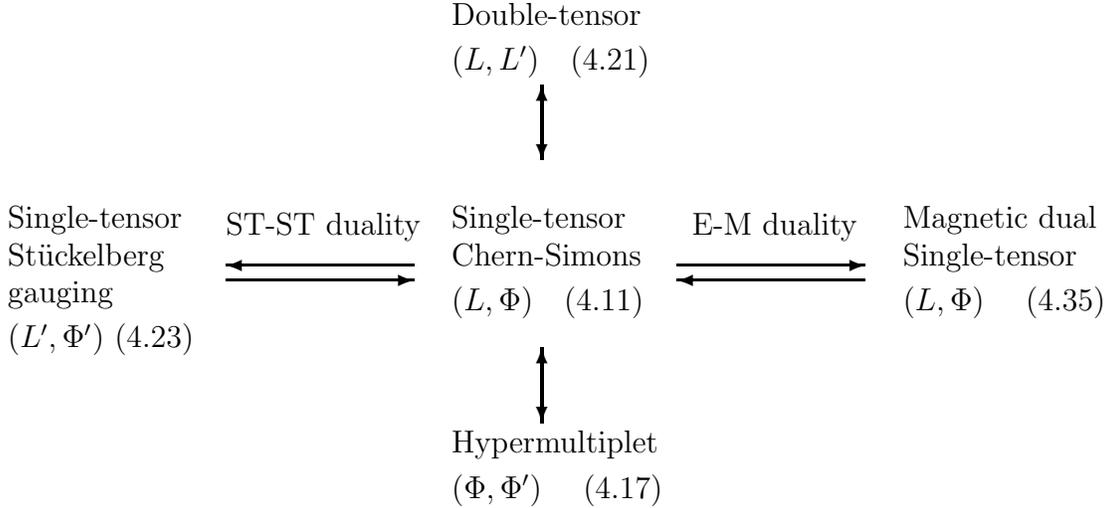

Specializing to the case of the universal dilaton hypermultiplet, we determined the action 
completely in the rigid limit, using the Heisenberg symmetry of perturbative string theory, 
up to an arbitrary constant 
parameter which, in the quaternion-K\"ahler case of $N=2$ supergravity, corresponds to the 
string one-loop correction \cite{oneloop}. An interesting open question is to realize this 
decoupling limit directly from the supergravity-coupled system. 

We have shown how the above system applies to the Higgs phase of $N=2$ nonlinear QED 
coupled to a charged hypermultiplet. Allowing a hypermultiplet mass scale and a Fayet-Iliopoulos 
term in the two-dimensional parameter space, the vacuum structure includes phases with broken and 
unbroken linear $N=1$ supersymmetry and/or $U(1)$ gauge symmetry.

It is interesting to note that in the Higgs phase the goldstino vector multiplet combines with the 
hypermultiplet to form a $N=1$ massive vector and a massless chiral superfield. This novel 
super-Higgs mechanism is possible without gravity because the hypermultiplet is charged under 
the $U(1)$ partner of the goldstino. In the $N=1$ case, the goldstino multiplet can be gauged only 
by gravity and is absorbed by the gravitino that acquires a mass.

In principle, it is straightforward to introduce additional hypermultiplets. Obviously only one of them 
will `absorb' the goldstino providing mass to the $U(1)$ vector.
This action describes also the low-energy limit of spontaneous partial supersymmetry breaking 
$N=2\to N=1$, when the breaking is `small' in the matter (hypermultiplet) sector. This is analogous, 
in the case of a single $N=1$ nonlinear supersymmetry, to the effective action of the goldstino 
coupled to $N=1$ multiplets at energies higher than their soft breaking masses. It is then known 
that this action is obtained by simply identifying the constrained goldstino multiplet with the so-called 
spurion~\cite{KS}. One may try to develop the analogy in the $N=2$ nonlinear 
case and derive the structure of possible `soft' terms associated to the partial $N=2\to N=1$ 
breaking. As a step further, one could try to integrate out the $N=2$ superpartners and obtain the 
effective action at much lower energies, describing the interactions of the $N=2$ goldstino multiplet 
to $N=1$ superfields. This would be directly relevant for constructing brane effective theories involving 
non-abelian gauge groups and charged matter. It could also be used for studying a supersymmetric 
extension of the Standard Model in the presence of a second supersymmetry nonlinearly realized 
due to its breaking at a high scale.

%*****************************************************
\section{Acknowledgements}
%*****************************************************
We wish to thank L. Alvarez-Gaum\'e and S. Ferrara for useful conversations. 
This work has been supported by the Swiss National Science Foundation. The work of
J.-P. D. was in part supported by CNRS and the Ecole Polytechnique, Palaiseau.
The work of I.A. and P.T. was supported in part by the European Commission under the ERC Advanced Grant 226371 and the contract PITN-GA-2009-237920. I.A. was also supported by the CNRS grant GRC APIC PICS 3747. P. T. is also supported by the ``Propondis" Foundation.

\renewcommand{\thesection}{\Alph{section}}
\setcounter{section}{0}
%********************************************************
\section{Conventions for \boldmath{$N=1$} superspace}\label{A1}
\setcounter{equation}{0}
%********************************************************
\renewcommand{\theequation}{\thesection.\arabic{equation}}

The $N=1$ supersymmetry variation of a superfield $V$ is
$\delta V = ( \epsilon Q + \ov\epsilon\ov Q )V$, with supercharges verifying the algebra
\beq
\label{conv3}
\{ÊQ_\alpha, \ov Q_\dalpha\} =  -2i (\sigma^\mu)_{\alpha\dalpha} \, \partial_\mu .
\eeq
On $V$, the supersymmetry algebra is then
\beq
\label{conv4}
[ \delta_1 , \delta_2 ] V = -2i \, ( \epsilon_1\sigma^\mu\ov\epsilon_2 
- \epsilon_2\sigma^\mu\ov\epsilon_1 ) \, \partial_\mu V.
\eeq
The covariant derivatives
\beq
\label{conv5}
D_\alpha = \frac{\partial}{\partial \theta^\alpha} - i(\sigma^\mu\ov\theta)_\alpha
\, \partial_\mu \, , 
\qquad\qquad
\ov D_\dalpha = \frac{\partial}{\partial \ov\theta^\dalpha} 
- i(\theta\sigma^\mu)_\dalpha \, \partial_\mu 
\eeq
anticommute with supercharges and verify
\beq
\label{conv6}
\{ÊD_\alpha, \ov D_\dalpha\} =  -2i (\sigma^\mu)_{\alpha\dalpha} \, \partial_\mu 
\eeq
as well. The identities
\beq
DD\,\theta\theta = \ov{DD}\,\ov{\theta\theta} = -4, 
\qquad\qquad
\Dint = -\frac{1}{4}\Fint \ov{DD} = -\frac{1}{4}\Fbarint DD,
\eeq
only valid under a space-time integral $\int d^4x$, are commonly used. 

The $N=1$ supersymmetry variations of the components $(z,\psi,f)$ of a chiral superfield 
$\Phi$, $\ov D_\dalpha\Phi=0$, are
\beq
\label{conv12}
\begin{array}{rcl}
\delta z &=& \sqrt 2 \, \epsilon\psi \, , 
\crbig
\delta \psi_\alpha &=& -\sqrt2 \, [ f \epsilon_\alpha  
+ i(\sigma^\mu\ov\epsilon)_\alpha \partial_\mu z] \, ,
\crbig
\delta f &=& -\sqrt2 \, i \, \partial_\mu\psi\sigma^\mu\ov\epsilon.
\end{array}
\eeq
The bosonic expansions of the chiral superfields used in the text are:
\beq
\label{convexp}
\begin{array}{rcl} 
W_\alpha(y,\theta) &=& \theta_\alpha d(y)
+{i\over 2}(\theta\sigma^\mu\ov{\sigma}^\nu)_\alpha F_{\mu\nu}(y) ,
\crbig
\chi_\alpha(y,\theta) &=& -{1\over 4}\theta_\alpha C(y)
+{1\over 4}(\theta\sigma^\mu\ov{\sigma}^\nu)_\alpha\, b_{\mu\nu}(y),
\crbig
\Phi(y,\theta) &=& \phi(y)-\theta\theta f_\phi(y),
\end{array}
\eeq
and any other chiral superfield has an expansion similar to $\Phi$. In this notation $\ov{\chi}_{\dalpha}=(\chi_\alpha)^*$ but $\ov{W}_\dalpha=-(W_\alpha)^*$. Since 
$L = D^\alpha\chi_\alpha - \ov D_\dalpha\ov\chi^\dalpha$, the linear superfield has
bosonic expansion 
\beq
\label{conv7}
\begin{array}{l}
L(x,\theta,\ov\theta) = C + \theta\sigma^\mu\ov\theta v_\mu + 
{1\over4}\theta\theta\ov{\theta\theta} \, \Box C, 
\crbig \hspace{2.9cm}
v_\mu = {1\over 2}\epsilon_{\mu\nu\rho\sigma}\partial^\nu b^{\rho\sigma}
= {1\over 2}\epsilon_{\mu\nu\rho\sigma}\partial^{[\nu}b^{\rho\sigma]}
= {1\over6} \epsilon_{\mu\nu\rho\sigma}H^{\nu\rho\sigma}.
\end{array}
\eeq
With these expansions, 
$$
\Dint \left[- L^2 + {1\over2}(\Phi+\ov\Phi)^2 \right]
$$
is the Lagrangian of a free, canonically-normalized, single-tensor $N=2$ multiplet. Its bosonic
content is
$$
{1\over2} (\partial_\mu C)(\partial^\mu C) + {1\over12} H_{\mu\nu\rho}H^{\mu\nu\rho},
\qquad\qquad
H_{\mu\nu\rho} = 3\, \partial_{[\mu}b_{\nu\rho]}.
$$

These identities are useful:
$$
\begin{array}{ll}
D_\alpha D_\beta = \frac{1}{2}\epsilon_{\alpha\beta} DD ,
\qquad\qquad
&\ov D_\dalpha \ov D_\dbeta = -\frac{1}{2}\epsilon_{\dalpha\dbeta}\ov{DD} ,
\crbig
[ D_\alpha , \ov{DD} ] = -4i(\sigma^\mu\ov D)_\alpha \partial_\mu ,
\qquad\qquad
&[ \ov D_\dalpha , DD ] = +4i(D\sigma^\mu)_\dalpha \partial_\mu ,
\crbig
DD\, W_\alpha = 4i(\sigma^\mu\partial_\mu\ov W)_\alpha ,
\qquad\qquad
&\ov{DD}\, \ov W_\dalpha = - 4i(\partial_\mu W\sigma^\mu)_\dalpha   .
\end{array}
$$
Further identities (with identical conventions) can be found in an appendix of Ref.~\cite{ADM}.

%*****************************************
\section{Solving the quadratic constraint}
\setcounter{equation}{0}
%*****************************************

The quadratic constraint ${\cal Z}^2 =0$ must be solved to obtain the magnetic DBI theory 
coupled to a single-tensor multiplet. Using the expansion
$$
{\cal Z}(y,\theta,\tilde\theta) = Z(y,\theta) + \sqrt2\, \tilde\theta \omega(y,\theta)
- \tilde\theta\tilde\theta \left[ {i\over2}\Phi_{\cal Z} + {1\over4}\ov{DD} \ov Z(y,\theta) \right],
$$
in terms of the $N=1$ chiral superfields $Z$, $\omega_\alpha$ and $\Phi_{\cal Z}$, the constraint is
equivalent to the single equation
\beq
\label{AppB1}
Z = - { \omega\omega \over i\Phi_{\cal Z} + {1\over2}\ov{DD}\ov Z} .
\eeq
The electric constraint equation (\ref{DBI7}), which was solved by Bagger and Galperin
\cite{BG} using a method which applies to Eq.~(\ref{AppB1}) as well, corresponds to the particular 
case $\omega_\alpha= iW_\alpha$, $\Phi_{\cal Z} = -i/\kappa$ and $Z=X$. Following then
Ref.~\cite{BG}, the solution of Eq.~(\ref{AppB1}) is
\beq
\label{AppB2}
Z(\omega\omega,\Phi_{\cal Z}) = 
{i\over \Phi_{\cal Z}}\left( \omega\omega + \ov{DD} \left[ { \omega\omega \ov{\omega\omega} \over 
|\Phi_{\cal Z}|^2 + A + \sqrt{|\Phi_{\cal Z}|^4 +2A|\Phi_{\cal Z}|^2 + B^2}} \right]\right),
\eeq
where
$$
\begin{array}{rcl}
A &=& -{1\over2}(DD\,\omega\omega + \ov{DD}\,\ov{\omega\omega}) \,\, =\,\, A^*,
\crbig
B &=& -{1\over2}(DD\,\omega\omega - \ov{DD}\,\ov{\omega\omega})\,\, =\,\, -B^*.
\end{array}
$$
Another useful expression is
\beq
\label{AppB3}
\begin{array}{l}
Z(\omega\omega,\Phi_{\cal Z}) = \displaystyle
{i\over \Phi_{\cal Z}} \Biggl( \omega\omega 
\crbig   \hspace{1.5cm}
\displaystyle + \ov{DD} \left[ 
{ \omega\omega \ov{\omega\omega} \over (DD\omega\omega) (\ov{DD}\ov{\omega\omega})}
\Bigl\{ |\Phi_{\cal Z}|^2 + A - \sqrt{|\Phi_{\cal Z}|^4 +2A|\Phi_{\cal Z}|^2 + B^2}\Bigr\}  \right] \Biggr).
\end{array}
\eeq
In the text, we need the bosonic content of $Z(\omega\omega,\Phi_{\cal Z})$. We write:
\beq
\label{AppB4}
\omega_\alpha (y,\theta)= \theta_\alpha\, \rho + {1\over2}(\theta\sigma^\mu\ov\sigma^\nu)_\alpha 
P_{\mu\nu} + \dots,
\eeq
where $\rho$ is a complex scalar (2 bosons), $P_{\mu\nu}$ a real antisymmetric tensor (6 bosons)
and dots indicate omitted fermionic terms. Hence,
$$
\begin{array}{rcl}
\omega\omega &=& \theta\theta \left[ \rho^2 + {1\over2} P^{\mu\nu} P_{\mu\nu}  
+ {i\over4}\epsilon^{\mu\nu\rho\sigma}P_{\mu\nu} P_{\rho\sigma} \right] + \dots,
\crbig
A &=& 2(\rho^2 + \ov \rho^2) + 2 P^{\mu\nu} P_{\mu\nu} + \dots,
\crbig
B &=& 2(\rho^2 - \ov \rho^2) + i \epsilon^{\mu\nu\rho\sigma}P_{\mu\nu} P_{\rho\sigma} + \dots
\end{array}
$$
Since the bosonic expansion of $\omega_\alpha$ carries one $\theta_\alpha$, it follows from 
solution (\ref{AppB2}) that the 
bosonic $Z(\omega\omega,\Phi_{\cal Z})$ has a $\theta\theta$ component only, and that this 
component only depends on $\rho$, $P_{\mu\nu}$ and the lowest scalar component of 
$\Phi_{\cal Z}$ (which we also denote by $\Phi_{\cal Z}$). As a consequence, the bosonic 
$Z(\omega\omega,\Phi_{\cal Z})$ does not depend on the auxiliary scalar $f_{\Phi_{\cal Z}}$ of 
$\Phi_{\cal Z}$. We then find:
\beq
\label{AppB5}
Z(\Phi_{\cal Z}, \omega\omega)_{bos.} = 
{i\ov\Phi_{\cal Z}\over |\Phi_{\cal Z}|^2}\omega\omega - {i\ov\Phi_{\cal Z}\over4|\Phi_{\cal Z}|^2} 
\theta\theta\left(|\Phi_{\cal Z}|^2 + A - \sqrt{|\Phi_{\cal Z}|^4 +2A|\Phi_{\cal Z}|^2 + B^2} 
\right)_{\theta=0}. 
\eeq
The parenthesis is real. In terms of component fields:
\beq
\label{AppB6}
\begin{array}{rcl}
Z &=& - {i \ov\Phi_{\cal Z} \over 4|\Phi_{\cal Z}|^2} \theta\theta \Big[ |\Phi_{\cal Z}|^2 
- i\epsilon^{\mu\nu\rho\sigma}P_{\mu\nu} P_{\rho\sigma} - 2 (\rho^2-\ov \rho^2)\Big]
\crbig
& & + {i \ov\Phi_{\cal Z} \over 4|\Phi_{\cal Z}|^2} \theta\theta\Bigl[ \Bigl(|\Phi_{\cal Z}|^2 
+ 2(\rho^2+\ov \rho^2)\Bigr)^2 - 16 \rho^2 \ov \rho^2 
+ 4 (\rho^2-\ov \rho^2)i\epsilon^{\mu\nu\rho\sigma}P_{\mu\nu} P_{\rho\sigma}
\crbig
& & \hspace{28mm} + 4 |\Phi_{\cal Z}|^2P^{\mu\nu} P_{\mu\nu} 
- \Bigl(\epsilon^{\mu\nu\rho\sigma}P_{\mu\nu} P_{\rho\sigma}\Bigr)^2 \Big]^{1/2} + \dots
\end{array}
\eeq
The decomposition (\ref{DBIme}), ${\cal Z}= \widetilde{\cal W} + 2g {\cal Y}$, indicates that
\beq
\label{AppB7}
\rho=-{g\over 2}C+i \widetilde d_2, 
\qquad P_{\mu\nu}= g b_{\mu\nu}- \widetilde F_{\mu\nu}\,,
\qquad\Phi_{\cal Z}=2g\Phi.
\eeq
In Lagrangian (\ref{DBIml}), we need the imaginary part of the $\theta\theta$ component of
$Z(\omega\omega,\Phi_{\cal Z})$:
\beq
\label{AppB8}
\begin{array}{rcl}
\Im Z(\omega\omega,\Phi_{\cal Z})|_{\theta\theta} &=& -{g\Re\Phi\over2} + {\Re\Phi\over8g|\Phi|^2}
\Biggl\{16g^4|\Phi|^4 + 8g^2|\Phi|^2(g^2C^2-4\tilde d_2^2) -16g^2C^2\tilde d_2^2
\crbig
&& + 16g^2 |\Phi|^2 (\widetilde F_{\mu\nu} - g \, b_{\mu\nu})(\widetilde F^{\mu\nu} - g \, b^{\mu\nu})
\crbig
&& + 8gC\tilde d_2 \, \epsilon^{\mu\nu\rho\sigma}
(\widetilde F_{\mu\nu} - g \, b_{\mu\nu})(\widetilde F_{\rho\sigma} - g \, b_{\rho\sigma})
\crbig
&&
- \Bigl[\epsilon^{\mu\nu\rho\sigma}
(\widetilde F_{\mu\nu} - g \, b_{\mu\nu})(\widetilde F_{\rho\sigma} - g \, b_{\rho\sigma}) \Bigr]^2
\Biggr\}^{1/2}
\crbig
&& + {\Im\Phi\over8g|\Phi|^2} \left[ \epsilon^{\mu\nu\rho\sigma}
(\widetilde F_{\mu\nu} - g \, b_{\mu\nu})(\widetilde F_{\rho\sigma} - g \, b_{\rho\sigma})
- 4gC \widetilde d_2 \right].
\end{array}
\eeq
We now use 
\beq
\label{AppB9}
\begin{array}{rcl}
-{\rm det} (|\Phi|\eta_{\mu\nu} + {\sqrt2\over g}\,P_{\mu\nu} ) 
&=& - |\Phi|^4 \,{\rm det} (\eta_{\mu\nu} + {\sqrt2\over g|\Phi|}\, P_{\mu\nu} )
\crbig
&=& |\Phi|^4 +{|\Phi|^2\over g^2}P^{\mu\nu}P_{\mu\nu} - 
{1\over16g^4}(\epsilon^{\mu\nu\rho\sigma}P_{\mu\nu}P_{\rho\sigma})^2 
\end{array}
\eeq
to rewrite
\beq
\label{AppB10}
\begin{array}{rcl}
\Im Z(\omega\omega,\Phi_{\cal Z})|_{\theta\theta} &=& -{g\Re\Phi\over2} + {\Re\Phi\over4g|\Phi|^2}
\Biggl\{ - 4g^4|\Phi|^4 \,{\rm det} \left[\eta_{\mu\nu} - {\sqrt2 \over g|\Phi|}(\widetilde F_{\mu\nu} 
-gb_{\mu\nu}) \right]
\crbig
&&
- 4 g^2 \tilde d_2^2 \Bigl(2|\Phi|^2 + C^2 \Bigr) + 2 g^4 C^2 |\Phi|^2
\crbig
&& + 2gC\tilde d_2 \, \epsilon^{\mu\nu\rho\sigma}
(\widetilde F_{\mu\nu} - g \, b_{\mu\nu})(\widetilde F_{\rho\sigma} - g \, b_{\rho\sigma})
\Biggr\}^{1/2}
\crbig
&& + {\Im\Phi\over8g|\Phi|^2} \left[ \epsilon^{\mu\nu\rho\sigma}
(\widetilde F_{\mu\nu} - g \, b_{\mu\nu})(\widetilde F_{\rho\sigma} - g \, b_{\rho\sigma})
- 4gC \widetilde d_2 \right].
\end{array}
\eeq
As a check, choosing $\Phi = -1/(2g\kappa)$ and $g=0$ to decouple the single-tensor multiplet leads back to theory (\ref{DBI8}) since in that case $\tilde d_2=0$.

\newpage

%\bibliographystyle{JHEP}
%\bibliography{references}

\end{document}